\documentclass[11pt,preprint]{aastex}
\usepackage{placeins}

\shortauthors{BAE ET AL.}
\shorttitle{EVOLUTION OF PHOTOEVAPORATING DISKS}

\begin{document}

\title{LONG-TERM EVOLUTION OF PHOTOEVAPORATING PROTOPLANETARY DISKS}

\author{Jaehan Bae\altaffilmark{1},
Lee Hartmann\altaffilmark{1},
Zhaohuan Zhu\altaffilmark{2},
Charles Gammie \altaffilmark{3,4}}

\altaffiltext{1}{Dept. of Astronomy, University of Michigan, 500
Church St., Ann Arbor, MI 48105} 
\altaffiltext{2}{Department of Astrophysical Sciences, Princeton University,
4 Ivy Lane, Peyton Hall, Princeton, NJ 08544}
\altaffiltext{3}{Dept. of
Astronomy, University of Illinois Urbana-Champaign, 1002 W. Green
St., Urbana, IL 61801}
\altaffiltext{4}{Department of Physics, University of Illinois Urbana-Champaign, 1110 W. Green St., Urbana, IL 61801}

\email{jaehbae@umich.edu, lhartm@umich.edu, zhuzh@astro.princeton.edu, gammie@illinois.edu}

\newcommand\msun{M_{\odot}}
\newcommand\lsun{L_{\odot}}
\newcommand\msunyr{M_{\odot}\,{\rm yr}^{-1}}
\newcommand\be{\begin{equation}}
\newcommand\en{\end{equation}}
\newcommand\cm{\rm cm}
\newcommand\mum{\mu \rm m}
\newcommand\kms{\rm{km~s^{-1}}}
\newcommand\K{\rm K}
\newcommand\etal{{et al}.\ }
\newcommand\sd{\partial}
\newcommand\mdot{\dot{M}}
\newcommand\rsun{R_{\odot}}
\newcommand\yr{\rm yr}

\begin{abstract}

We perform calculations of our one-dimensional, two-zone disk model to study the long-term evolution of the circumstellar disk.
In particular, we adopt published photoevaporation prescriptions and examine whether the  photoevaporative loss alone, coupled with a range of initial angular momenta of the protostellar cloud, can explain the observed decline of the frequency of optically-thick dusty disks with increasing age.
In the parameter space we explore, disks have accreting and/or non-accreting transitional phases lasting of $\lesssim20~\%$ of their lifetime, which is in reasonable agreement with observed statistics.
Assuming that photoevaporation controls disk clearing, we find that initial angular momentum distribution of clouds needs to be weighted in favor of slowly rotating protostellar cloud cores.  Again, assuming inner disk dispersal by photoevaporation, we conjecture that this skewed angular momentum distribution is a result of fragmentation into binary or multiple stellar systems in rapidly-rotating cores.  
Accreting and non-accreting transitional disks show different evolutionary paths on the $\dot{M}-R_{\rm wall}$ plane, which possibly explains the different observed properties between the two populations.
However, we further find that scaling the photoevaporation rates downward by a factor of 10 makes it difficult to clear the disks on the observed timescales, showing that the precise value of the photoevaporative loss is crucial to setting the clearing times.
While our results apply only to pure photoevaporative loss (plus disk accretion), there may be implications for models in which planets clear disks preferentially at radii of order 10 AU.

\end{abstract}


\section{INTRODUCTION}

Studies over the last 20 years have shown that optically-thick dusty disks around
young stars have a significant range in
their lifetimes, with a median value of $\sim3$~Myr \citep{haisch01,hernandez07,mamajek09}. 
While the dust emission of these disks may effectively disappear via
accretion onto the central star or coagulation into larger bodies including
planetesimals and planets, it is difficult to remove disk gas and dust on
radial scales larger than a few tens of AU via accretion on short enough timescales
\citep{hollenbach00}
and there are challenges even forming gap-producing giant planets at a few AU on
these lifetimes \citep[see][and references therein]{lissauer07}.

The currently most popular mechanism for removing outer disk material is photoevaporation by
stellar high-energy radiation - either ultraviolet or X-rays, or both - from the central star \citep{clarke01,alexander06,ercolano09,gortiandhol09,gorti09,owen10,owen11,owen12}.\footnote{We exclude photoevaporation in clusters by massive stars; see e.g. \citet{hollenbach00}.}
Photoevaporative loss can also produce gaps and inner holes in disks, potentially explaining disks with large optically-thin holes inside of optically thick outer regions - the so-called transitional
disks \citep{calvet05} - as well as pre-transitional disks with large gaps but with optically-thick 
dust in the innermost regions \citep{espaillat07}.
Alternatively, giant planets can clear disk gaps \citep{dodson11}, though there may be difficulties in clearing large enough disk regions \citep{z11}, and Type II migration may also be a problem \citep{clarke13}.

Observationally, (pre-)transitional disks are usually identified via either detailed modeling of 
the spectral energy distributions (SEDs) \citep[e.g.,][]{calvet02,calvet05,espaillat07,espaillat08,kim13} or high-resolution imaging \citep[e.g.,][]{hughes07,hughes09,andrews09,andrews10,andrews11}.
The gaps and inner holes in (pre-)transitional disks are optically thin but may not be completely
evacuated of dust and gas \citep[e.g.,][]{rosenfeld12,kraus13}, especially as many of these objects
exhibit gas accretion onto the central star \citep[e.g.,][]{calvet02,calvet05}.
Dust growth and/or filtration is probably required to explain many of the observations 
\citep{dullemond05, rice06,z12b}.  It should be emphasized that both SED modeling and current submillimeter interferometric imaging are not able to identify very small disk gaps and inner holes.

As mentioned above, optically-thick pre-main sequence disks exhibit a variety of lifetimes,
with the frequency of such disks steadily declining from ages $\sim 1$~Myr to $\sim 10$~Myr.
\citet{alexander09} could reproduce the observed disk frequencies by including migrating,
gap-opening planets with a range of initial masses at 5 AU while varying the initial disk masses.
\citet{owen11} were also able to explain the variation of disk lifetimes
by assuming that similar stars have very different X-ray fluxes, as seen in observations of,
for example, the deep X-ray observations of the Orion Nebula Cluster \citep{Preibisch05}.
However, it is known that X-ray emission from very young low-mass stars is highly time-variable 
on timescales of days or less due to flares \citep{montmerle83,feigelson02a,caramazza07}, as well as
on timescales of years, the mechanisms for which are uncertain \citep{feigelson02b,feigelson03}.
The level of variability of X-ray emission over relevant timescales for disk evolution of $10^4$ to $10^6$~yr
are of course unknown, but is arguably smaller than the spread observed at a given epoch.

An alternative hypothesis, which we explore here, is that disk lifetimes mostly depend upon the 
initial angular momentum of the protostellar cloud.  As we show, more rapid rotation
leads to larger disks which take longer to deplete.  
Even if X-ray luminosity for individual stars
exhibits a wide dispersion over long timescales, substantial differences in initial angular momenta are to be expected and their effects should not be ignored.
In addition, much of the previous work on disk photoevaporation has been performed under the
assumption of fully-viscous disks; this is questionable for low-temperature
protostellar/protoplanetary disks, where ionization states can be far too low for the
magnetorotational instability (MRI) to operate - the only mechanism so far known to act in
a manner similar to viscous transport \citep[e.g.,][and references therein]{balbus98}.
The presence of ``dead zones'' \citep{gammie96}, i.e. regions in which angular momentum transport is very limited or does not occur,
can strongly affect disk structure, and by inference the manner in which photoevaporative
mass loss changes that structure.

In this paper we further develop the two-zone disk model of \citet{z10a,z10b}, modified as in 
\citet[][Paper I]{bae13} to allow for non-zero dead zone residual viscosity, by adding a prescription for 
photoevaporation taken from the calculations of \citet{owen12}.  
We follow the disk evolution from the time of formation during the protostellar phase and analyze the subsequent disk structure during the later, T Tauri phase.
As the actual amount of photoevaporative flux (or equivalently, the mass-loss rate it drives) is uncertain, we adopt the fiducial value of \citet{owen12} and test 30 and $10~\%$ of that rate.  
We explore the distribution of initial angular momenta that would be required to explain disk frequencies, and we compare on the predictions of the photoevaporation $+$ accretion model
with observations.
We find that the lower photoevaporation rate is insufficient to explain the observations, but the fiducial level is adequate, coupled with a frequency distribution of initial cloud angular momenta that is strongly skewed toward low values.  We suggest that this skewed distribution is not necessarily representative of protostellar core angular momenta distributions, but rather reflects fragmentation at higher core rotation.

\section{METHODS}
\label{sec:methods}

We use the one-dimensional, two-zone disk model introduced in \citet{z10a,z10b} and modified in Paper I, including photoevaporation by energetic photons from the central star
\citep{ercolano09,owen10,owen11,owen12}.
Using their radiation-hydrodynamic models, \citet{owen11,owen12} calculate and provide fits to the total mass-loss rate due to X-ray photoevaporation as a function of stellar mass and X-ray luminosity  as follows.
\begin{eqnarray}
\dot{M}_{PE} & = & 6.25\times10^{-9}\left(\frac{M_*}{1~M_\odot}\right)^{-0.068}\left(\frac{L_X}{10^{30}~{\rm erg~s^{-1}}}\right)^{1.14}~M_{\odot}~{\rm yr^{-1}}~{\rm for~full~disks}
\label{eqn:mdot_lx_p}
\\
\dot{M}_{PE} & = & 4.8\times10^{-9}\left(\frac{M_*}{1~M_\odot}\right)^{-0.148}\left(\frac{L_X}{10^{30}~{\rm erg~s^{-1}}}\right)^{1.14}~M_{\odot}~{\rm yr^{-1}}~{\rm for~disks~with~inner~holes}
\label{eqn:mdot_lx_t}
\end{eqnarray}
We note that the photoevaporative mass-loss rate scales nearly linearly with X-ray luminosity and depend weakly on stellar mass.

\begin{figure}
\begin{center}
\includegraphics[scale=0.8]{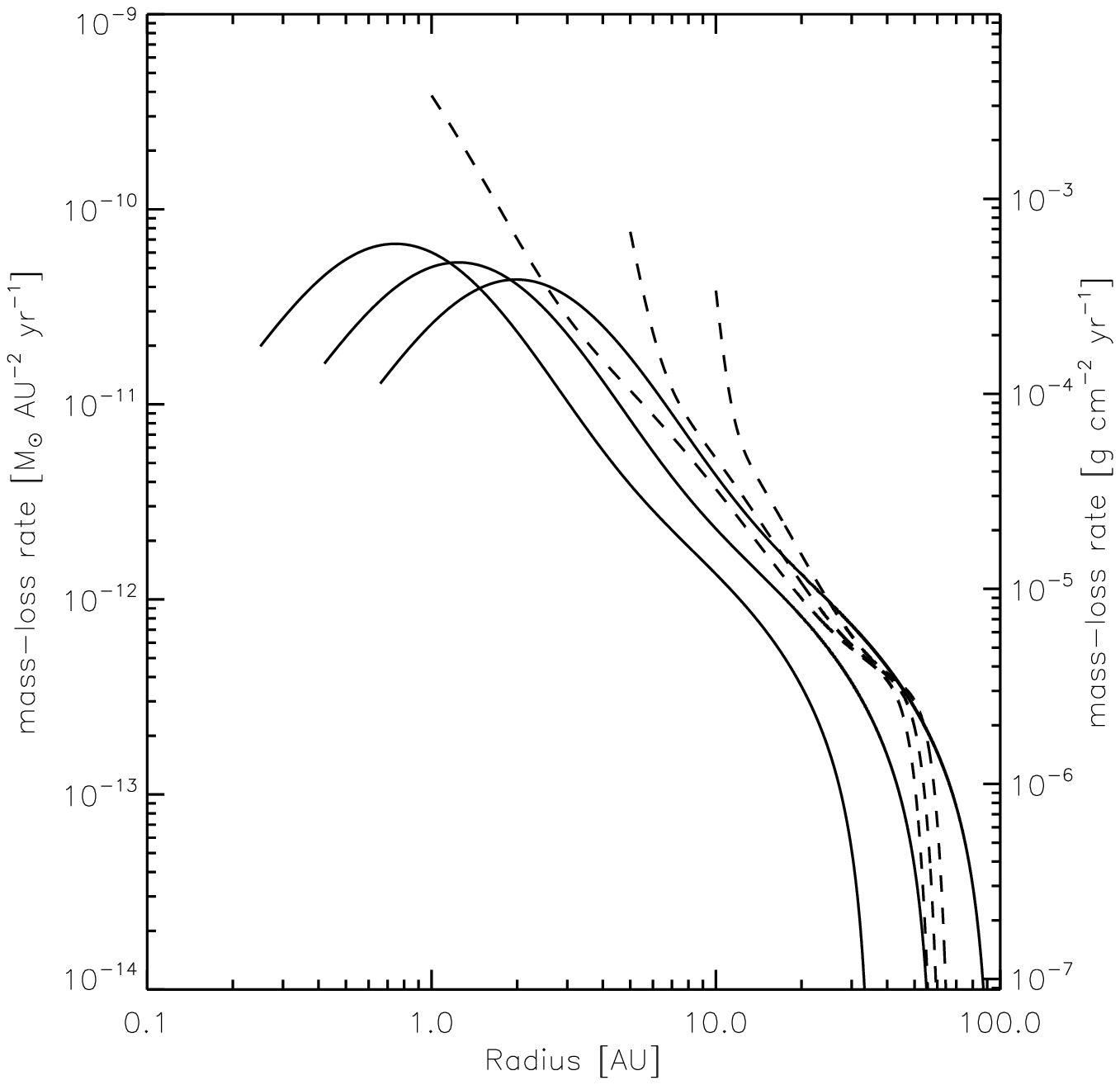}
\caption{Radial mass-loss profiles due to photoevaporation. Solid curves present mass-loss profiles of full disks for three different stellar masses, $M_*=$~0.3, 0.5, and 0.8~$M_\odot$ (from left to right). High energy photons most efficiently evaporate disk material at small radii ($R \lesssim 3$~AU) and the mass-loss becomes negligible beyond $\sim100$~AU. Dashed curves show mass-loss profiles of disks with inner holes at $R=1, 5$, and 10~AU (from left to right), assuming a $0.8~\msun$ central star. Compared to the primordial disk with the same central stellar mass, photoevaporation rate at the inner edge of the disk is enhanced.} 
\label{fig:PE_model}
\end{center}
\end{figure}

\citet{Preibisch05} studied nearly 600 X-ray sources that can be reliably identified with optically well-characterized T Tauri stars in the Orion Nebula Cluster.
They found that low-mass ($M_* \le 2~M_{\odot}$) stars show a clear correlation between X-ray luminosity and stellar mass as
\begin{equation}
\log(L_X~[{\rm erg~s^{-1}}])=30.37(\pm0.06)+1.44(\pm0.10)\log(M_*/M_{\odot}).
\label{eqn:lx_mass}
\end{equation}
By combining Equations (\ref{eqn:mdot_lx_p}), (\ref{eqn:mdot_lx_t}), and (\ref{eqn:lx_mass}), we obtain 
the
total photoevaporation rate in terms of stellar mass only.
\begin{eqnarray}
\dot{M}_{PE}=1.65\times10^{-8}(M_*/M_{\odot})^{1.57}~M_{\odot}~{\rm yr^{-1}}~{\rm for~full~disks}
\label{eqn:mdot_pe_p}
\\
\dot{M}_{PE}=1.27\times10^{-8}(M_*/M_{\odot})^{1.49}~M_{\odot}~{\rm yr^{-1}}~{\rm for~disks~with~inner~holes}
\label{eqn:mdot_pe_t}
\end{eqnarray}
Using the total mass-loss rate and the normalized radial mass-loss profile provided by \citet[][see their Appendix B]{owen12}, we calculate a scaled mass-loss profile $\dot{\Sigma}_{PE}$ for a given stellar mass that satisfies $\dot{M}_{PE} = \int{2\pi R \dot{\Sigma}_{PE}}~dR$.
The resulting radial mass-loss profiles for 0.3, 0.5, and 0.8~$\msun$ stars are presented in Figure \ref{fig:PE_model}.

The photoevaporation term is added in the mass conservation equation as
\be
\label{eqn:mass_eqn}
2\pi R {\partial \Sigma_i \over \partial t} - {\partial \dot{M}_i \over \partial R} = 2\pi g_i (R, t) - 2\pi R {\dot{\Sigma}_{PE}}
\en
and this equation is solved together with the angular momentum conservation equation as explained in Paper I.
In Equation (\ref{eqn:mass_eqn}), $\Sigma_i$ is the surface density, $\dot{M}_i$ is the radial mass flux, and $2\pi g_i (R,t)$ is the mass flux per unit distance of infall material from an envelope cloud.
The subscript $i$ denotes either the active layer (``a") or dead zone (``d").
Since it is unlikely that photons travel through the dusty infalling material, we turn on photoevaporation after infall ends.
We thus note that the stellar mass and the corresponding photoevaporation rate do not change dramatically during the photoevaporative period.

As shown in Figure 1, the photoevaporative mass-loss rate is enhanced when there is an inner hole,
as the inner edge of the optically-thick disk is illuminated normally rather than obliquely.  
We turned on the photoevaporative rate of Equation (\ref{eqn:mdot_pe_t}) when the inner disk is
essentially totally depleted.  In principle, once a large gap is opened and dead zone is absent, there should be an enhancement
of the photoevaporative rate which gradually grows as the shadowing by the inner disk decreases
over time; but we find that the clearing time of the inner disk is so short compared with the
overall disk lifetime that it makes little difference when exactly we turn on inner edge
photoevaporation.

As in \citet{z10b} and in Paper I, we follow the disk evolution during and after infall from
a protostellar cloud which forms the star as well as the disk.  
We assume that the protostellar cloud is spherical with a two-component density profile 
intended to approximate a Bonnor-Ebert sphere as explained in \citet{z10b}; 
the flat core inside of $r_{ic}$ has $0.1~\msun$ and the rest of the Bonnor-Ebert sphere ($r > r_{ic}$) has $1~\msun$, with a density profile of $r^{-2}$.  
We assume that the inner core material
collapses first to form a central star, and then follow the evolution after that event.
The outer cloud beyond $r_{ic}$ collapses as in the \citet{terebey84} model for an initially-uniformly rotating singular isothermal sphere,
with infall landing at progressively large radii as the collapse proceeds. 
If the cloud core is initially in uniform rotation with angular velocity $\Omega_c$, the material falling in from different directions will have different angular momentum and arrive at the midplane at different radii within the centrifugal radius \citep{cassen81},
\be
\label{eqn:rc}
R_c = r_0^4 {\Omega_c^2 \over G M_c}.
\en
Here, $M_c$ is the central stellar mass and $r_0$ is the initial radius of material that collapses to the center at time $t$, defined as $r_0 = (m_0/2)c_s t + r_{ic}$, where $m_0=0.975$ \citep{shu77,terebey84}, $c_s$ is the (uniform) sound speed of the cloud, and $r_{ic}$ is the radius of the flat density region in the Bonnor-Ebert sphere.
As seen in Equation (\ref{eqn:rc}), for a fixed cloud temperature the size of protoplanetary disks are mainly determined by the initial angular momentum of their maternal cloud.
Infall material is added within the centrifugal radius following the modified infall model explained in Paper I.

We parameterize the cloud rotation in terms of $\omega = \Omega_c / \Omega_{\rm b}$, where $\Omega_{\rm b} = 2^{3/2} c_s^3/G M_{\rm c}$ is the breakup angular frequency at the outer cloud edge.
We choose $\omega$ in a way that the maximum centrifugal radius at the end of infall spans from 1 to 500~AU, which corresponds to $0.007 \le \omega \le 0.1$.
Our choice of $\omega$ and the corresponding maximum centrifugal radii are presented in Table \ref{table:results_wope}.

Active layer and dead zone temperatures are determined by the balance between heating and radiative cooling.
As explained in Paper I, for the heating sources, we consider local heating by the viscosity, the infall, and the gravitational potential energy change, and external heating by stellar and accretion luminosity irradiations.
The Rosseland mean opacity was taken from \citet{zhu09} to calculate the optical depths of the layers.

Following Paper I, we consider a dead zone residual viscosity (DZRV) $\alpha_{\rm rd}$ as well as the MRI viscosity and the GI viscosity; $\alpha_d  = \alpha_{M,d} + \alpha_{Q,d} + \alpha_{\rm rd}$.
In non-zero DZRV models, $\alpha_{\rm rd}$ is set to 
\be
\label{eqn:alpha}
\alpha_{\rm rd} = {\rm min} \left( 10^{-4},~f_{\rm rd} \alpha_{\rm MRI} {\Sigma_a \over \Sigma_d}\right),
\en
where $f_{\rm rd}$ is the efficiency of accretion in the dead zone and set to unity.
This form is motivated by the conjecture that the accretion driven in the dead zone by the active layer MRI cannot be larger than that of the active zone itself (Paper I).
The MRI viscosity parameter $\alpha_{\rm MRI}$ is a fixed value of which a region have when it can sustain the MRI.
In the dead zone we turn the MRI on and set $\alpha_{M,d}=\alpha_{\rm MRI}$ only if the midplane temperature becomes higher than a critical temperature $T_{\rm MRI}$ to produce sufficient ionization levels, while the active layer is always assumed to be able to sustain the MRI.
The GI viscosity parameter is the same as in \citet{z10b}, $\alpha_{Q,i}=e^{-Q^2}$, where $Q$ is the Toomre parameter.

\section{RESULTS}
\label{sec:results}

\subsection{Initial Conditions}

In all cases here we start with a $0.1~M_{\odot}$ central protostar surrounded by an $1~M_{\odot}$ cloud.
Our fiducial models assume a cloud rotation of $\omega = 0.03$ and we vary $\omega$ in other models as explained in \S\ref{sec:methods}.
We assume $T_{\rm MRI}=1500$~K and $\alpha_{\rm MRI}=0.01$ for all calculations.
We adopt a cloud envelope temperature of $T_{\rm env}=20$~K, 
which yields a constant infall rate of $\sim3.4\times10^{-6}~\msunyr$.
This is $20~\%$ smaller than the infall rate for
conventional singular isothermal collapse model \citep{shu77} because of the modified 
infall model (see Paper I).
The infall lasts for $\sim 0.24$~Myr, adding $0.8~M_{\odot}$ to the central star + disk in total.

We use inner and outer disk boundaries of 0.2~AU and 8000~AU, respectively, with 64 logarithmically spaced grid points.  
These values result in the innermost grid size of 0.036~AU.
We tested our results using differing grid cells after implementing photoevaporation prescriptions and found that the results are not sensitive to the resolution.
We adopt outflow boundary conditions for both inner and outer boundaries.
This assumes that all the material passing the inner boundary to be accreted onto the central star.
At the outer boundary, the amount of material leaving is essentially zero so that the disk evolution is independent of the outer boundary conditions.

\subsection{Evolution of Non-photoevaporating Disks}
\label{sec:non-photoevaporating_disks}

In  Paper I we presented the evolution of non-photoevaporating disks up to ages of 1.5~Myr, 
focusing on the effects of a non-zero residual viscosity in dead zone.
These models showed a variety of patterns of non-steady accretion onto the central star,
especially during the phase when matter from the protostellar envelope is still being
added to the disk.  The reasons for the outburst behavior and their parameter dependence
were discussed in Paper I.  Here we focus on the longer-term evolution of the disks
when outbursts are either absent or relatively unimportant.

\begin{figure}
\begin{center}
\includegraphics[scale=1]{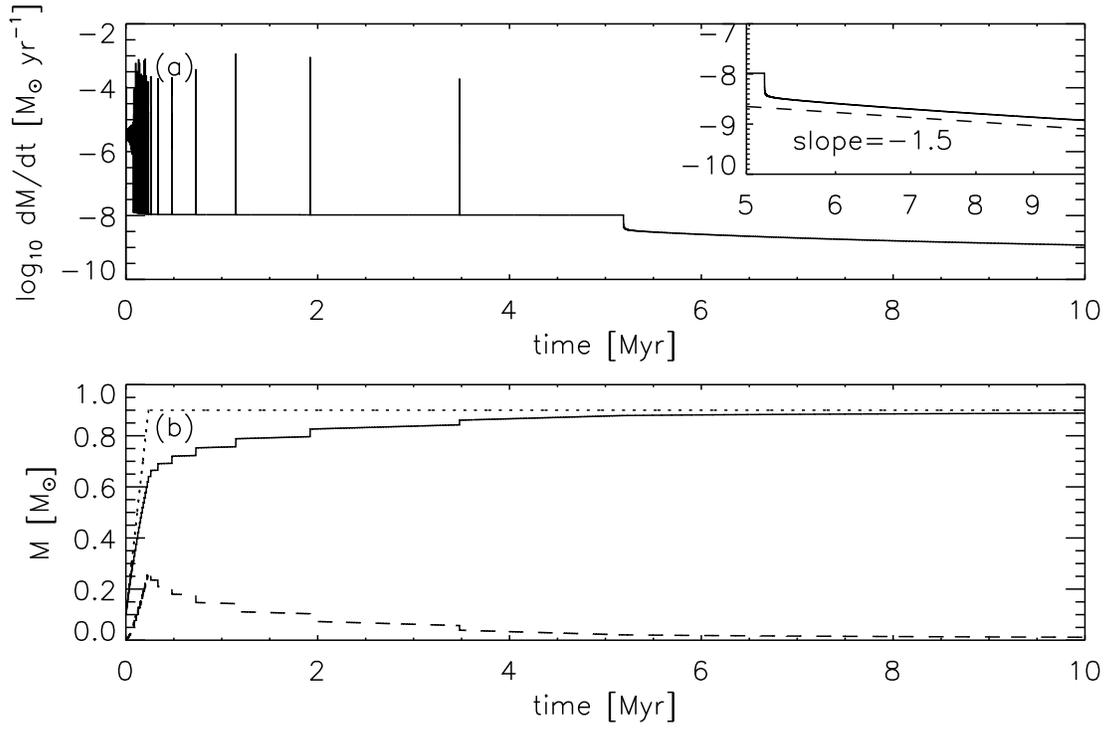}
\caption{(a) Mass accretion rate and (b) mass of the central star + disk (dotted curve), mass of the central star (solid curve), and mass of the disk (dashed curve) with time for our fiducial non-photoevaporating disk with zero DZRV. The upper right panel in (a) shows mass accretion rate vs. log(time) after the dead zone depletes and the disk becomes fully viscous. The dashed line presents the similarity solution for fully viscous disks, $\dot{M} \propto t^{-3/2}$ \citep{hartmann98}.}
\label{fig:evol_z_wope}
\end{center}
\end{figure}

Figure \ref{fig:evol_z_wope} presents the mass accretion rate and the mass of the central star, the disk, and the central star + disk of our fiducial zero DZRV model as a function of time.
As seen in Paper I, the system shows a quasi-steady disk accretion at the beginning, an outburst stage during infall, and the T Tauri phase after infall ends.
The dead zone depletes at $\sim5.2$~Myr and the disk becomes fully viscous, having only the active layer.
Mass accretion rate after the dead zone depletes is presented in the upper-right corner of Figure \ref{fig:evol_z_wope}a, together with the similarity solution for viscous disk evolution, $\dot{M} \propto t^{-3/2}$ \citep{hartmann98}.
Not surprisingly, mass accretion rate matches the similarity solution well.

\begin{figure}
\begin{center}
\includegraphics[scale=1]{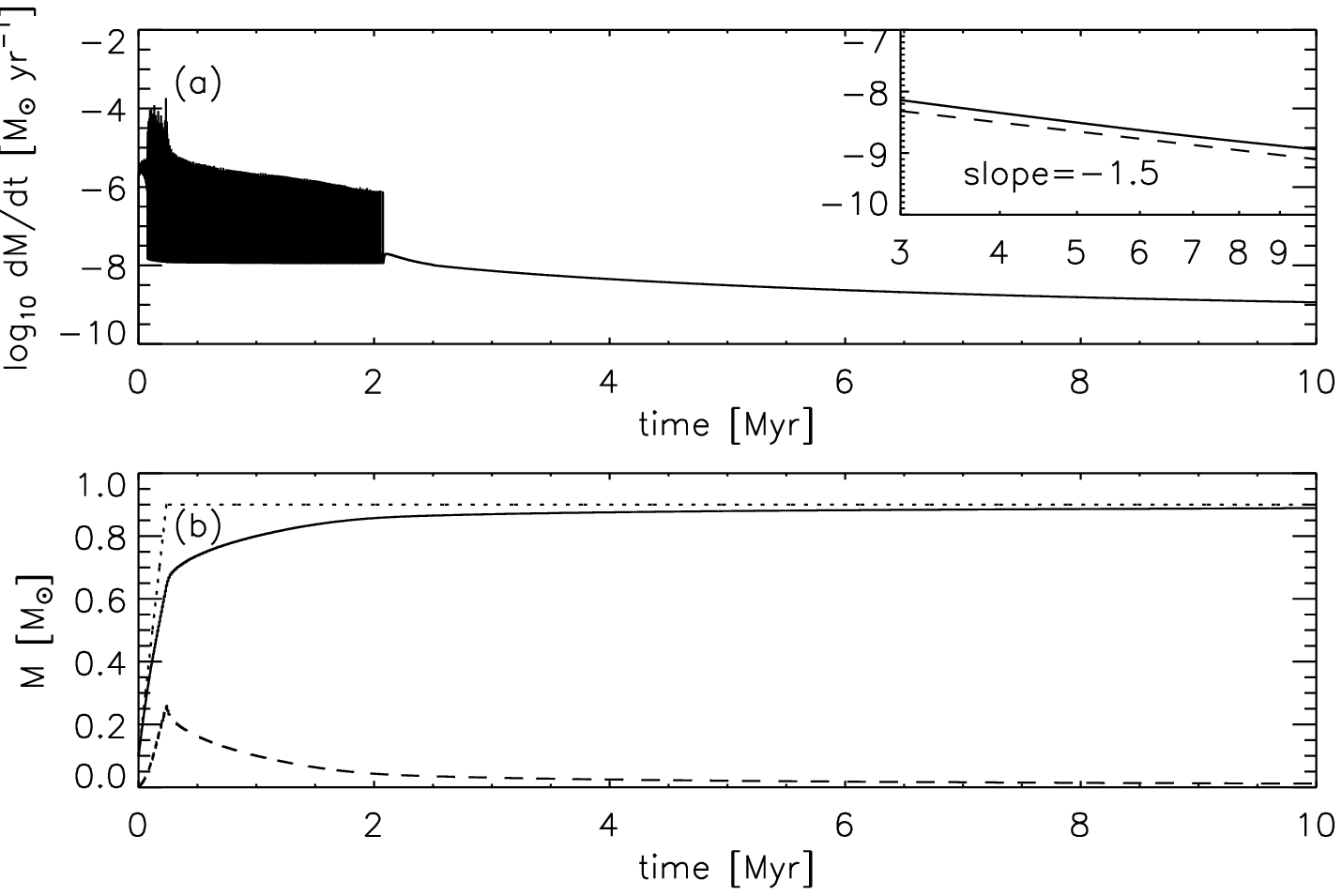}
\caption{Same as Figure \ref{fig:evol_z_wope} but results with non-zero DZRV.}
\label{fig:evol_nz_wope}
\end{center}
\end{figure}

In Figure \ref{fig:evol_nz_wope}, we present the mass accretion rate and the mass of the central star, the disk, and the central star + disk of the ficucial non-zero DZRV model.
In this case, the disk accretes material more efficiently through small but frequent inside-out bursts (see Paper I), resulting in earlier depletion of dead zone than zero DZRV model, at $\sim2.5$~Myr.
Again, the viscous evolution after dead zone depletes follows the similarity solution well.

\begin{figure}
\begin{center}
\includegraphics[scale=0.8]{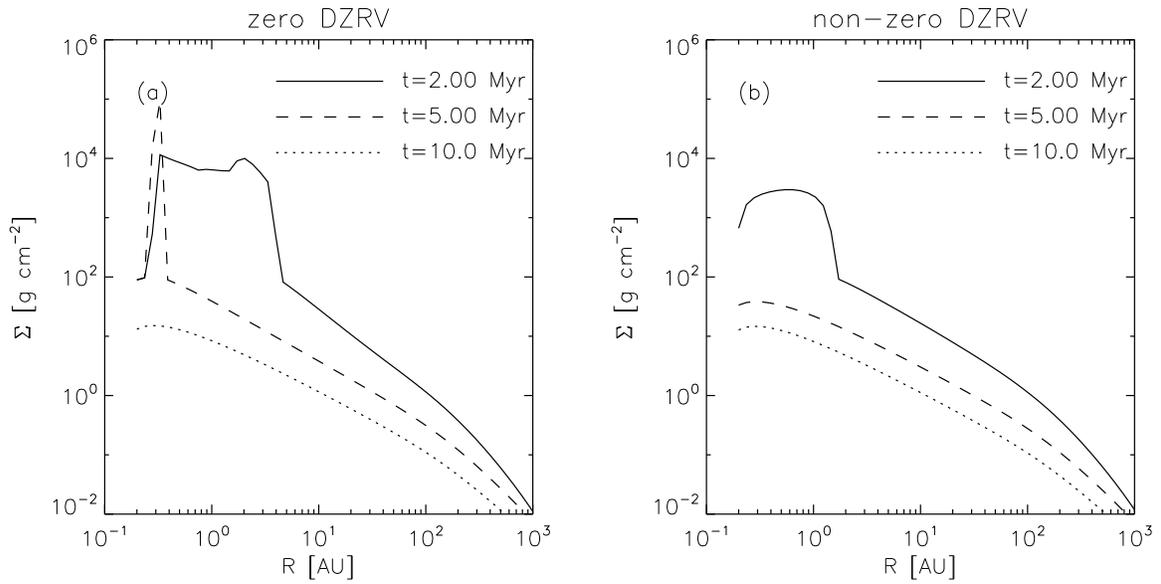}
\caption{Surface density profiles of the fiducial non-photoevaporating models with (a) zero and (b) non-zero DZRV, at  2~Myr (solid curves), 5~Myr (dashed curves), and 10~Myr (dotted curves). While density distribution at small radii ($R \lesssim 10$~AU) is largely affected by outbursts, the disk properties at large radii ($R \gtrsim 10$~AU) is controlled more by viscous evolution, showing similar profiles in both models.}
\label{fig:radialpl}
\end{center}
\end{figure}

Surface density distributions of both zero and non-zero DZRV models at 2, 5, and 10~Myr are presented in Figure \ref{fig:radialpl}.
As seen in the figure, outbursts govern density structures at small radii ($R \lesssim 10$~AU), so the detailed density profiles at such radii depend on the DZRV during outbursting phase.
However, the disk properties at large radii ($R \gtrsim 10$~AU) are likely to be controlled more by viscous evolution regardless of the DZRV, and thus the disk properties at such radii appear to be similar in both models.

\subsection{Evolution of Photoevaporating Disks}
\label{sec:photoevaporating_disks}

\begin{figure*}
\centering
\includegraphics[scale=1]{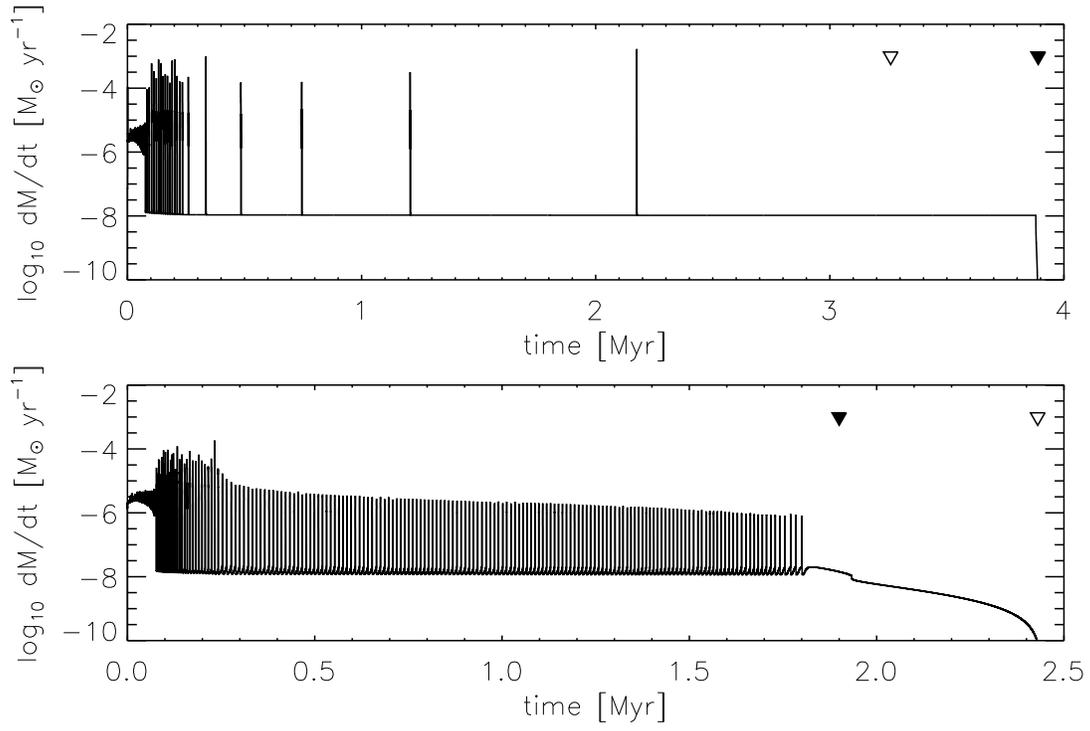}
\caption{Mass accretion rate of photoevaporating models with zero DZRV (upper) and non-zero DZRV (lower) as a function of time. Open triangles indicate the time when a gap is opened, while solid triangles present the time when dead zone is depleted. Note that a gap opens before the dead zone depletes in the zero DZRV case, while the dead zone depletes first and then a gap opens afterward in the non-zero DZRV case.}
\label{fig:evolpe}
\end{figure*}

Figure \ref{fig:evolpe} shows the mass accretion rates of the fiducial photoevaporating models with zero and non-zero DZRV.
The overall evolutions resemble those of non-photoevaporating disks.
However, accretion stops early at $\sim3.9$ and $\sim2.4$~Myr, as the inner disk depletes via accretion plus photoevaporation.
We note that a gap opens first and then dead zone depletes in the zero DZRV model, while dead zone depletes first and then a gap opens later in the non-zero DZRV model (see below).

Figure \ref{fig:radialpe}a presents the surface density distributions of the zero DZRV model at three times; when infall ends, a gap opens, and the wall temperature of outer disk drops below 100~K (see \S\ref{sec:trans_disk}).
When infall ends the disk has a maximum centrifugal radius of $R_{c,{\rm max}} \sim 25$~AU, which approximately sets the size of the dead zone \citep{z10b}.
Having no infalling material the dead zone loses mass purely through episodic accretion and shrinks in size, but the process is relatively inefficient due to zero DZRV.
Meanwhile, photoevaporative winds deplete the outer disk beyond the dead zone and open a gap around $R\sim9$~AU, at $\sim3.3$~Myr.
A gap can be opened beyond the dead zone because photoevaporation rate decreases approximately as $\sim R^{-2}$ at $5 \lesssim R \lesssim 50$~AU while disk surface density at the radii drops steeper.
The inner disk totally depletes at $\sim3.9$~Myr through accretion and photoevaporation.

Figure \ref{fig:radialpe}b shows the density distributions of the non-zero DZRV model.
The disk surface density when infall ends is similar to that of the zero DZRV model.
However, the dead zone depletes earlier at $\sim1.9$~Myr with the help of frequent inside-out bursts, not allowing photoevaporative winds enough time to make a gap beyond the dead zone.
As a result, a gap opens after surface density drops significantly through episodic accretion, around $R\sim1$~AU.
Since the gap opens at small radii, the inner disk has little mass within it so depletes immediately after the gap opens.
After the inner disk depletes, the outer disk is cleared out rapidly due to the enhanced photoevaporative winds at the inner edge.

\subsection{Effect of Initial Angular Momentum}
\label{sec:am}

As explained in \S\ref{sec:methods} the disk size is mainly constrained by the centrifugal radii, or initial angular momentum.
\citet{z10b} showed that if a system has small initial angular momentum it results in a small disk with little mass, having only a few (or no) outbursts.
On the other hand, if a system starts with large angular momentum disk becomes larger and more massive, having numerous outbursts.
We confirm that the same holds in our non-photoevaporating models.
The results are summarized in Table \ref{table:results_wope}.

By adding photoevaporation and varying initial angular momenta, we find that disks have gaps and inner holes during their evolution.
Figure \ref{fig:evolpew_z} shows the evolution of photoevaporating disks with small ($\omega=0.012$) and large ($\omega=0.055$) initial angular momenta, assuming zero DZRV.
With the small initial angular momentum, plenty of material is accreted onto the central star during infall and only $0.09~\msun$ of material is left in the disk when infall ends.
In addition, the maximum centrifugal radius is small enough ($R_{c,{\rm max}}=3.1$~AU) so that high energy photons are able to evaporate disk material and open a gap beyond the dead zone.

With the large initial angular momentum, on the other hand, the disk has a large maximum centrifugal radius of $R_{c,{\rm max}} = 108.2$~AU, resulting in a massive disk of $0.40~\msun$ at the end of infall.
In this model the dead zone extends out to $\sim50$~AU, constrained by the radius where the active layer becomes gravitationally unstable rather than the maximum centrifugal radius \citep{z10b}.
A significant fraction of the disk mass resides in the large radii and therefore photoevaporation cannot open a gap until the disk is sufficiently depleted.
A gap finally opens at 4.74~Myr after the dead zone depletes, around 2~AU.

We show the evolution of photoevaporating disks with non-zero DZRV in Figure \ref{fig:evolpew_nz}.
In the small initial angular momentum case dead zone depletes early at 1.17~Myr with the help of frequent inside-out bursts.
A gap opens near the inner disk edge and evolves immediately into an inner hole.
In the large initial angular momentum case disks have plenty of mass at large radii, so a gap cannot be opened until the disk depletes sufficiently.
We note again that once an inner hole is created the inside-out disk clearing occurs in a short timescale compared to the disk lifetime.
The results of the photoevaporating models are summarized in Table \ref{table:results_wpe}.

\subsection{Comparison with Other Studies}
\label{sec:comp}

Our results bear some qualitative similarities to the results of \citet[][hereafter OEC11]{owen11}, 
where the same photoevaporation profile was used, but also some discrepancies.
In OEC11, they assumed a $0.7~\msun$ central star with photoevaporation of $7.1\times10^{-9}~\msunyr$, and the zero-time similarity solution of \citet{lynden-bell74} with disk mass of $0.07~\msun$, characteristic disk radius of 18~AU, and $\alpha=2.5\times10^{-3}$.
In terms of masses of the central star and the disk, their model is close to our model with $\omega=0.012$; when infall ends the model has a $0.8~\msun$ central star with a $0.09~\msun$ surrounding disk.

In order to compare our models to OEC11's, we test the $\omega=0.012$ model with the same photoevaporation rate as OEC11, which is $\sim60~\%$ of our fiducial rate.
As presented in Figure \ref{fig:radialpel}, with our layered disk model we can have either a gap beyond dead zone or an inner hole depending on the DZRV.
The final outcome of OEC11 is similar to our non-zero DZRV model in that a gap opens at similar radii and the outer disk clears out rapidly once an inner hole is created (see their Figure 9).
However, in their model the gap opens at $\sim2.7$~Myr, during which photoevaporative winds are capable of opening a gap beyond dead zone in our case.
In Figure \ref{fig:radialpel} we overplot initial surface density distribution of OEC11.
While their zero-time similarity solution is close to our density distribution in general, it has about an order of magnitude higher surface density at $10 \lesssim R \lesssim 50$~AU where we open a gap in our zero DZRV model.
Due to the high initial surface density at these radii, a gap opens close to the disk inner edge after surface density drops sufficiently through the viscous evolution.
To summarize, despite the qualitative resemblance between models, this comparison shows how sensitive the disk evolution is to details of the model one uses.

\citet{morishima12} used a layered accretion model (active layer + dead zone) adopting a similar photoevaporation prescription to ours and tested for various parameter sets of X-ray luminosity, active layer surface density, and dead zone viscosity parameter.\footnote{Dead zone viscosity in \citet{morishima12} is the total viscosity in the dead zone, different from our dead zone ``residual'' viscosity.}
While details are different, results of \citet{morishima12} are in qualitative agreement with ours; for a given X-ray luminosity gaps open beyond the dead zone if the dead zone has small viscosity ($\alpha_d \leq 10^{-5}$ with their standard X-ray luminosity) and thus lives long, while gaps open at a small radius only after the dead zone disappears if dead zone viscosity is large enough ($\alpha_d \geq 10^{-4}$) for the mass accretion in the dead zone to be efficient (see their Figure 4).

\subsection{Efficiency of Photoevaporation}
\label{sec:efficiency}

Since the actual amount of mass-loss due to photoevaporation is uncertain we examined cases  with $30~\%$ and $10~\%$ of the standard rate.
Figure \ref{fig:evolpel} presents the mass accretion rates of the models as a function of time.
With $30~\%$ less photoevaporation rate, neither zero DZRV model nor non-zero DZRV model opens a gap beyond dead zone.
With $10~\%$ of the standard photoevaporation rate, the mass accretion rates show no significant difference from those of non-photoevaporating disks.
The only difference is that photoevaporative winds help to deplete the inner disk and make an inner hole.
The inner hole in the models is created at 9.57 and 9.31~Myr, respectively, leaving a $0.01~\msun$ outer disk.

\section{COMPARISON WITH OBSERVATIONS}

While an increasing number of (pre-)transitional disks have been observed, the 
causes of the disk gaps and inner holes are still uncertain.
Theories suggested to explain (pre-)transitional disks include photoevaporation \citep{alexander07,gortiandhol09,gorti09,owen10,owen11,owen12}, planet formation \citep{dodson11,z11}, grain growth \citep{dullemond05}, etc.
In the following we investigate the implications of the models assuming that photoevaporation is the only mechanism for disk clearing.

\subsection{Photoevaporation-driven Transitional Disks}
\label{sec:trans_disk}

We showed in \S\ref{sec:results} that different combinations of parameters result in a variety of evolutionary results, including disks with gaps and inner holes.
Those disks obtained in our calculations could be potential analogues of observed (pre-)transitional disks.
Observationally, the SED of (pre-)transitional disks are characterized by the near-IR flux close to photospheric emission and the rise of mid-IR flux around $\sim10~\mum$.
We assume that this excess emission can be observed only if gaps and inner holes are large enough; $\Delta R_{\rm  gap} > 5$~AU and $R_{\rm hole}>1$~AU \citep{strom89,calvet05,espaillat07,espaillat10,luhman12}.
In principle we could define pre-transitional disks as disks with gaps that are large enough to satisfy the above criteria and 
thus distinguishable from transitional disks.
However, since our models do not include dust evolution, it is uncertain whether or not there will be observable dust emission 
from the inner disk.
Thus, we do not distinguish between pre-transitional disks and transitional disks in this work.
Instead, we classify the disks into ``accreting" and ``non-accreting" transitional disks depending on the existence of accretion onto the central star (i.e. existence of an inner disk regardless of the presence of dusts).

With the definitions above and assuming the standard photoevaporation rate, accreting transitional disks can be obtained with zero DZRV plus small initial angular momentum ($\omega \lesssim 0.04$), while non-accreting transitional disks require either non-zero DZRV or zero DZRV with a large initial angular momentum ($\omega > 0.04$).
Figure \ref{fig:phase} shows various disk phases in the disk age vs.\ $\omega$ space.
In the zero DZRV cases with initial angular frequency of $\omega \lesssim 0.04$, the maximum centrifugal radius is small enough so that high energy photons are able to evaporate disk material and open a gap beyond dead zone, as explained in \S\ref{sec:photoevaporating_disks}.
When $\omega>0.04$, on the other hand, the disk has a large centrifugal radius so gaps cannot be opened until the inner disk is sufficiently depleted and thus disks end up with a non-accreting transitional phase.
In the non-zero DZRV cases, all disks in our $\omega$ space evolve into non-accreting transitional disks due to a short dead zone depletion time.

We assume that transitional disks become evolved when the outer ``wall'' -- the inner edge of the outer disk -- has a 
temperature less than 100~K (we assume a blackbody temperature for simplicity) with no inner disk left so that we don't 
expect to see excess emission at $8~\mu{\rm m}$ (the longest wavelength channel of the sensitive IRAC camera on the {\em Spitzer Space Telescope}).
In our models, the wall temperature drops below 100~K at $R_{\rm wall}\sim20$~AU.
The assumption is based on the fact that most of the previous surveys employed the IR excess criteria to study protoplanetary census.
With this definition, transitional disks in our models have outer disk clearing time of $\lesssim20~\%$ of their lifetime, which reasonably explains the observed statistics \citep{luhman10,muzerolle10}.

\subsection{Disk Frequency}

The disk fraction of young stellar objects is known to decrease as a function of their age.
\citet{mamajek09} compiled results from the literature and showed that the 
disk frequency can be fitted approximately by an exponential decay with a characteristic time scale of $\sim2.5$~Myr. 
Here we try to explain the observed disk frequency using photoevaporating disks with different initial angular momenta.

Since we know more rapid rotation leads to a longer disk depletion time, we can calculate the disk frequency as follows.
If we write the distribution of initial angular momentum of systems as $n(\omega)$, the disk frequency at a given time can be written as 
\begin{eqnarray}
f(t) = {{1} \over {N_{\rm tot}}}{{\int_{\omega_{\rm evol}}^{\omega_{\rm max}}{n(\omega)}d\omega}},
\label{eqn:disk_freq}
\end{eqnarray}
where $N_{\rm tot} \equiv {\int_{\omega_{\rm min}}^{\omega_{\rm max}}{n(\omega)} d\omega}$ is the total number of systems, $\omega_{\rm min}$ and $\omega_{\rm max}$ are the minimum and the maximum initial angular velocities, and $\omega_{\rm evol}$ is the initial angular velocity of the disks turning into the evolved phase at time $t$, respectively.
Here we set $\omega_{\rm min}=0.007$ and $\omega_{\rm max}=0.1$ (see \S\ref{sec:methods}) because the point is we are trying to derive the initial angular momentum distribution from the disk frequency.
Thus, we take $f(t) = \exp(-t/\tau_{\rm disk})$ where $\tau_{\rm disk}=2.5$~Myr and estimate the initial angular momentum distribution.
By taking derivative of Equation (\ref{eqn:disk_freq}), 
\be
{{df} \over {dt}} = -{{1} \over {N_{\rm tot}}} {{d\omega_{\rm evol}} \over {dt}} n(\omega).
\label{eqn:am_dist}
\en
Then, since we have $d\omega_{\rm evol}/dt$ from our calculations (a derivative of the fit in Figure \ref{fig:phase}) we can estimate $n(\omega)$.

Figure \ref{fig:am_dist} presents the resulting cumulative distribution of initial angular momentum.
In both zero and non-zero DZRV cases about a half of the disks have initial angular momentum smaller than our fiducial value 0.03.
The distribution steeply decreases toward the populations with large initial angular momentum. 
In Figure \ref{fig:disk_freq}, we present disk frequency obtained assuming the initial angular momentum distribution we have in Figure \ref{fig:am_dist}.
The lack of initially rapidly-rotating systems might reflect initial conditions; however, a more plausible explanation is that rapidly-rotating protostellar cores end up fragmenting into binaries or multiple stellar systems \citep{kratter08,vorobyov10,z12a}.
In terms of specific angular momentum in units of ${\rm cm^2~s^{-1}}$, our models range from $\log j \sim 17.2$ to 20.3 at the end of infall and have less after infall ends as systems lose angular momentum through photoevaporative winds.
This is about two orders of magnitude smaller than the observed specific angular momenta of molecular clouds, $\log j \sim 19.6-22.2$ in units of ${\rm cm^2~s^{-1}}$ \citep{goodman93,barranco98, caselli02,andrews10}.
However, since it is questionable to directly compare angular momenta of disk systems to those of molecular clouds given observational difficulties \citep[e.g.][]{isella09}, here we limit comparison of our disks to evolved binary systems only.
If we take a median orbital period of nearby solar-like binaries $P\sim180$~yr \citep{duquennoy91}, a binary system with two $0.5~\msun$ stars has a specific angular momentum of $\log j \sim 20.4$ which is close to the highest end of the angular momentum distribution obtained in our models.
This may support the idea that rapidly rotating molecular cloud cores preferentially fragment into binaries or multiple stellar systems.

\subsection{Two Populations on the $\dot{M}-R_{\rm wall}$ Plane}
\label{sec:mdot_radii}

Detailed modeling of the broadband SEDs of individual disks enables us to search for gaps or inner holes in their density distribution and to estimate their sizes \citep[e.g.][]{calvet02,calvet05,espaillat07,espaillat08,espaillat10,espaillat12,kim13}.
Combined with mass accretion rate, these information may provide an important clue to the inner disk clearing mechanism. 

In Figure \ref{fig:mdot_rwall} we show the probability distribution of our transitional disks on the $\dot{M}-R_{\rm wall}$ plane.
We calculate the probability based on the time a disk spends at a given grid on the plane.
Then, we weight the probability with the initial angular momentum distribution estimated in the previous section.
We note that this does not change the overall trend significantly.
We additionally assume equal chance to have either zero DZRV or non-zero DZRV for simplicity.
The minimum accretion rate is arbitrarily set to $10^{-11}~\msunyr$.
The gradation in color from the left to right seen in the plot is because of using logarithmic radial grids.
In this analysis, we assume that we can verify gaps and inner holes out to large enough radii instead of using the evolved disk criteria described in \S4.1 since the studies modeling the SEDs of individual disks we refer here use broadband SEDs out to $40~\mu m$ \citep{kim13} or even further out to sub-mm/mm \citep{espaillat10,espaillat12}.

It is evident from Figure \ref{fig:mdot_rwall} that there are two distinct populations which occupy different regions and have different evolutionary paths on the $\dot{M}-R_{\rm wall}$ plane as well.
Accreting transitional disks open gaps while the dead zone still survives, so can have accretion rates as high as $\sim10^{-8}~\msunyr$ and have gaps at relatively large radii ($R_{\rm wall} \gtrsim 6$~AU). 
On the other hand, non-accreting transitional disks have lower accretion rate of $\lesssim10^{-10}~\msunyr$ and only produce small inner holes ($R_{\rm wall} \lesssim 3$~AU).

Data taken from the literature \citep{natta06,najita07,gudel07,brown09,hughes09,hughes10,espaillat10,merin10,andrews11,espaillat11,kim13} are overplotted on Figure \ref{fig:mdot_rwall}.
While we can see some objects that lie close to the predicted region with high probabilities, a large fraction of the observed disks have moderate accretion rates of $10^{-10}< \dot{M} <10^{-8}~\msunyr$.
However, it should be noted that the estimated probability of observing such objects with moderate accretion rates is low ($< 1~\%$).
This implies that it is unlikely for the observed transitional disks with moderate accretion rate to be transiting from accreting phase to non-accreting phase, as far as its short timescale is considered.
Instead, we propose that those disks can be explained by varying $\Sigma_A$ \citep[e.g.,][]{morishima12}, $\alpha_{\rm MRI}$, stellar mass, etc., which are fixed in our calculations.
Varying those parameters is, however, beyond the scope of this paper.

\subsection{Millimeter Flux Densities of Transitional Disks}

In order to test properties of transitional disks at millimeter wavelengths we calculate flux densities at 1.3~mm.
The face-on flux density is calculated as 
\be
F_\nu = {1 \over d^2} \int{B_\nu (R) 2 \pi R}dR,
\en
assuming a distance of $d=140$~pc.
Here, $B_\nu (R) = B_\nu (T_a(R)) (1-e^{-\tau_a(R)}) + B_\nu (T_d(R)) (1-e^{-\tau_d(R)}) e^{-\tau_a(R)}$ where $B_\nu$ is the Planck function. 
We use temperature profiles that are self-consistently calculated in our code as explained in \S\ref{sec:methods}.
Optical depths of the active layer and the dead zone, $\tau_a$ and $\tau_d$, are obtained with the general opacity of $\kappa_\nu = 0.1\left( {\nu / 10^{12}~{\rm Hz}} \right)~{\rm cm^2~g^{-1}}$ \citep{beckwith90}.

Figure \ref{fig:mmflux} shows the evolution of transitional disks on the ${\rm millimeter~flux} - R_{\rm wall}$ plane.
We see that both accreting and non-accreting transitional disks evolve in a way that the flux density declines and $R_{\rm wall}$ increases, with more massive disks having higher millimeter flux densities in general.
In addition, both accreting and non-accreting transitional disks spend most of their time in the large-cavity region 
($R_{\rm wall} \gtrsim 15$~AU, \citealt{andrews11}) of the plane; 
accreting transitional disks open gaps at large radii with an inner dead zone, 
while in non-accreting transitional disks inner holes open at small radii but grow 
quickly because of the absence of the dead zone and the enhanced photoevaporation at the 
inner edge of the disk (see Figure \ref{fig:PE_model}).
We note that transitional disks with large initial angular momenta have a longer disk clearing time 
(both absolute time and relative time to the total disk lifetime) than those with small initial angular momenta, 
which possibly explains a high transitional disk fraction in the millimeter-bright disks 
\citep{andrews11}, although the current results may be affected by the observational bias toward detecting 
larger holes, which favors larger disks with larger masses.

We note that a sudden drop in flux density occurs for accreting transitional disks as the inner disk depletes.
This is because the outer disk is already optically thin when a gap opens and thus the inner disk contributes a comparable amount of millimeter emissions (see Figures \ref{fig:radialpe}, \ref{fig:evolpew_z}, and \ref{fig:evolpew_nz}).
The disk mass, however, does not change significantly during the drop since the inner disk has only small mass in it.
Due to the emissions from the inner disk, while accreting, the accreting transitional disks have systematically higher flux densities than the non-accreting transitional disks with the same initial angular momentum, which presumably explain the observed millimeter-bright transitional disks having considerable higher accretion rates \citep{owenc12}.

However, we find a number of observed millimeter-faint transitional disks with small cavities as well (see Figure 1 of \citealt{clarke13} and references therein).
Our models can explain those transitional disks with small initial angular momenta plus non-zero DZRV, but given again that the inner cavity in such models grows rapidly one may need other inner disk clearing mechanisms such as planet formation and/or planet migration.

\subsection{Viscous Evolution of the Outer Disk}

While we have mainly focused on disk properties inside of $\lesssim 100$~AU so far, sub-mm and mm interferometric observations of CO line emission have shown that some gaseous disks extend out to several hundred AU or even further \citep{pietu07,schaefer09,oberg10,andrews12}.
In order to compare our results to observations we use a simple prescription for the viscous evolution of accretion disks \citep{lynden-bell74,hartmann98}.
The model assumes a viscosity that is written as a power law in $R$ as $\nu \propto R^\gamma$ and has a surface density profile of 
\be
\Sigma = (2-\gamma) {M_d \over 2\pi R_d^2} \left( R \over R_d \right)^{-\gamma} \exp \left[ -\left( R \over R_d \right)^{2-\gamma}\right],
\en
where $R_d$ is a characteristic scaling radius and $M_d$ is the disk mass.
As we show earlier, the evolution of outer disks at $\gtrsim 100$~AU is not affected significantly by outbursts.
In addition, the mass-loss due to photoevaporation at such large radii is negligible.
Therefore, it seems reasonable to assume that outer disks evolve viscously.

Figure \ref{fig:mass_radius} shows the best fit characteristic radii of density profiles together with disk masses at given times.
Although disk structures at small radii may be complicated, we see that outer disks evolve in the same way that angular momentum is conserved, $M_d \propto R_d^{-1/2}$.
Disks with different initial angular momenta spread roughly perpendicular to the evolution path at each given time.
The viscous evolution seen in our models is consistent with the observed disk mass -- radius relation \citep[e.g.][]{andrews11}.

Photoevaporating disks have systematically smaller masses than the non-photoevaporating disks due to mass-loss, but there seems to be no significant difference in the viscous evolution of outer disks between the two populations.
However, we see photoevaporation leads depopulation of disks in the lower right corner of the plots, especially in the non-zero DZRV models where disks evolve faster.

\section{DISCUSSION}

Our simulations show that photoevaporative loss alone, coupled with a range of 
initial angular momenta of the prostellar cloud, can in principle explain the
observed disk frequencies, rather than assuming variations in long-term X-ray luminosities
as in \citet{owen12}.  
However, in both treatments a rather high photoevaporative rate
is required which amounts to $\sim 10^{-8}~\msunyr$, basically the same as the typical disk
accretion rate.  
While direct, quantitative observational estimates of the photoevaporative mass-loss rate are difficult to make, unfortunately, the ``low-velocity" wind component seen in forbidden lines of T Tauri stars \citep{hartigan95} may represent the photoevaporative loss from disks. 
Using the same radiation-hydrodynamic model as in \citet{owen10,owen11,owen12}, \citet{ercolano10} showed that both the luminosities and profiles of the low-velocity component of the [O~I] forbidden lines observed by \citet{hartigan95} can be successfully reproduced.
Their most successful model assumes a $0.7~\msun$ central star with $L_X = 2\times10^{30}~{\rm erg~s^{-1}}$, which corresponds to photoevaporation rate of $\sim1.4\times10^{-8}~\msunyr$.
However, the predicted line luminosities of the winds depend very sensitively on the assumed X-ray luminosity (see Table 1 of \citealt{ercolano10}).

The required total photoevaporation rate of $\sim 10^{-8}~\msunyr$ also matches well with the expected disk gas dispersal of the solar system.
In our model photoevaporation yields mass-loss of $\sim100~{\rm g~cm^{-2}}$ per million years at Jupiter's orbit.
Inferred from the required initial surface density of solids to form Jupiter at its radii and solar chemical compositions, $\sim700~{\rm g~cm^{-2}}$ of gas is thought to initially be present at Jupiter's orbit \citep[][and references therein]{pollack96} but to be depleted over the typical disk lifetime suggested by observations, $\sim3$~Myr \citep{haisch01,hernandez07,mamajek09}.
Thus, within a factor of two our photoevaporative mass-loss rate fits the minimum-mass solar nebula model.

In our models, the dead zone usually has a surface density of $\gtrsim 10^4~{\rm g~cm^{-2}}$.
Assuming a 0.8~$\msun$ central star, photoevaporation removes maximum $\sim400~{\rm g~cm}^{-2}$ of material from the circumstellar disk per 1~Myr.
This means that without significant accretion it is difficult to deplete the dead zone through photoevaporation only within the typical disk lifetime. 
In our zero DZRV models, where the outside-in bursts are not efficient and/or not frequent enough to deplete disk material, the dead zone survives long and thus it may act as a potential trap for planets not to migrate inward rapidly \citep{hasegawa11,hasegawa12}.
The long-lived dead zone also prevents the outer disk from rapid evaporation through the enhanced photoevaporative and might provide favorable conditions for giant planet formation.

While our photoevaporating disk models can explain observations of the disk frequency vs. age and the transitional disk frequency, it is worthwhile to note that our results are sensitive to the assumed parameters.
As we showed in \S\ref{sec:comp} details of disk evolution, such as when and where a gap opens and whether a disk evolves into an accreting or a non-accreting transitional disk, etc., are significantly dependent upon the density structure of the disk.
Also, there should be some regions in which viscous transportation of material is very limited (i.e. a dead zone).
In this sense, the assumption of fully-viscous disk assumed in much of protoplanetary disk simulations should be used with caution.
Furthermore, our results are sensitive to the photoevaporation rate.
As pointed out in \S\ref{sec:efficiency}, we do not see accreting transitional disks with $30~\%$ less photoevaporation rate, and $10~\%$ of the standard rate turns out to be insufficient to significantly affect disk evolution.
Thus, photoevaporation is a crucial factor in long-term disk evolution.

Recently, \citet{kim13} claimed that the gaps in transitional disks of their sample are likely due to the gravitational influence of Jovian planets or brown dwarfs orbiting within the gaps, and that X-ray photoevaporation is not the dominant mechanism for creating transitional disks.
They draw the latter part of their conclusion by comparing their data with the theoretical predictions from the X-ray photoevaporation model by \citet{owen12}.
However, one of the main assumptions in \citet{owen12} is questionable; profiles of transitional disks are self-similar in a way that if a gap opens at 1~AU around a $1~\msun$ it would open at 0.1~AU around a $0.1~\msun$ star.
This may be true for fully viscous disks obeying the similarity solution \citep{lynden-bell74}.
However, as explained in \S\ref{sec:methods} disk size is mainly determined by the centrifugal radius $R_c $, which is proportional to $M_c^{-1}$ if we assume a constant $\Omega_c$ and is proportional to $M_c^{-3}$ if we assume a constant $\omega$ for different stellar masses.
Thus, if this is the case less massive stars will have less massive but more extended surrounding disks, which is not self-similar to more massive stars.
Also, the existence of the dead zone will complicate the details of disk structures and their evolution.

In addition, the probability of detecting accreting transitional disks in the predicted region by  photoevaporation model of OEC11 (the shaded region of Figure~24 in \citealt{kim13}) is low ($\lesssim1~\%$, see Figure~13 of OEC11), while \citet{owen12} later suggested that there may be a rapid dynamical clearing mechanism of the outer disk of non-accreting transitional disks, so-called thermal sweeping, i.e. X-ray penetrates further in the disk midplane as the bound X-ray heated gas starts to flow predominantly vertically, alleviating the low probability.

We note that there are some limitations in our analysis so caution is needed when interpreting the results.
First, our photoevaporation prescription is obtained from the simulations using disks with an approximate power-law radial structure \citep{owen10,owen11,owen12}.
Thus, it is not obvious that the mass-loss rate and/or profile are applied in the same way to disks with more complicated radial structure, such as the dead zone.
The mass-loss rate can be insensitive to disk structure as \citet{owen12} stated in their results, but more thorough calculations are needed in order to deal with this issue.
Second, we estimate the disk frequency based on the existence of IR excess at $8~\mu m$ assuming a simple blackbody wall temperature.
This implicitly assumes that gas and dust are well mixed and evolve in the same manner, which is not necessarily the case as several studies have shown observationally \citep[e.g.][]{ingleby11,andrews12} and numerically \citep[e.g.][]{alexander07}.
Third, close binaries can also disrupt disks, although the analysis of  \citet{kraus12} suggests that this may not affect the disk frequency vs. age (see their Figure 3).
Fourth, while observations include a wide range of stars in mass, we only consider $0.1~\msun$ core $+$ $1~\msun$ cloud.
We expect different initial masses will result in a more diverse evolution.
Fifth, note that our definition and classification of transitional disks are limited by the neglect of possible dust migration and evolution, which
might increase the ratio of accreting to non-accreting transition disks \citet{alexander07}.
Our results also do not rule out giant planet formation as the dominant mechanism for disk clearing; as \citet{rosetti13} suggest, a combination of planet gap opening and photoevaporation may be particularly effective in producing large disk gaps. 

Finally, we note the importance of using a small inner boundary in simulations.
In our photoevaporating models, the non-accreting transitional phase starts with gaps opened at small radii of $0.6-3.3$~AU (see Table \ref{table:results_wpe}).
Once an inner hole is created photoevaporative winds clear the outer disk rapidly, within $\sim0.01$~Myr in the fastest case for example.
Therefore, it is worth emphasizing that simulations need to use a small inner boundary of $R\lesssim 1$~AU to capture the rapid inside-out disk clearing via photoevaporation.

\section{SUMMARY}

In this study, we present long-term evolution of protostellar disks using the one-dimensional, two-zone disk model introduced in \citet{z10a,z10b} and modified in Paper I.
We found that disks accrete material more efficiently when they have non-zero viscosity in the dead zone, and thus have a shorter lifetime.

By adding photoevaporation to our models and varying initial angular momentum, we were able to produce accreting and non-accreting transitional disks with reasonable timescale of $\lesssim 20~\%$ of the total lifetime which is in reasonable agreement with observations.
We found that disks having non-zero DZRV preferentially evolve into  non-accreting transitional disks due to a short disk depletion time, while evolution of disks with zero DZRV varies depending on the initial angular momentum.
Starting with small initial angular momentum of $\omega \lesssim 0.04$, zero DZRV models evolve into accreting transitional disks with high accretion rate of $\sim10^8~\msunyr$.
We note that relatively inefficient accretion in the dead zone help the disks to have the accreting transitional disk phase.
On the other hand, for initially rapidly rotating cases with $\omega > 0.04$, zero DZRV models evolve into non-accreting transitional phase since a significant fraction of the disk mass resides in the large radii and thus photoevaporation cannot open a gap at such radii.

The observed disk frequency could be explained using our photoevaporating disk models by varying initial angular momenta, as an alternative to the adoption of varying long-term stellar X-ray luminosities by \citet{owen11}.
Assuming that photoevaporation controls disk evolution, we found the initial angular momenta need to be distributed in a way that decreasing significantly toward the systems with large initial angular momentum.
Comparison of angular momenta of our models to the observed nearby solar-type binaries  suggests that rapidly-rotating molecular cloud cores end up fragmenting into multiple systems with smaller disk, while those with slow rotation results in a single star-disk system.
Our models also explain the different observed properties between accreting and non-accreting transitional disks, each of which occupies different regions and follows different evolutionary paths on the $\dot{M}-R_{\rm wall}$ plane.

\acknowledgments

We acknowledge support from NASA grant NNX11AK53G and the use of the University of Michigan Flux cluster.
C.G. was supported in part by a Richard and Margaret Romano Professorial Scholarship.

%******************************************************

% figure 1

% figure 2

% figure 3

% figure 4

% figure 5

% figure 6
\begin{figure*}
\centering
\includegraphics[scale=0.8]{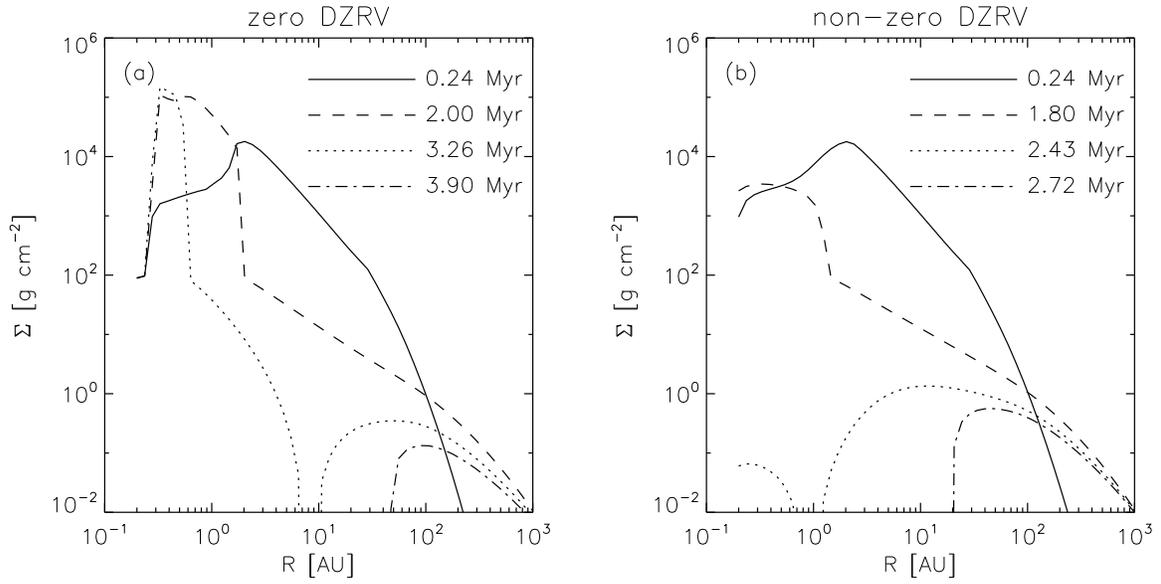}
\caption{Surface density profiles of photoevaporating models with (a) zero and (b) non-zero DZRV models at selected times. We note that dead zone survives long in the zero DZRV model, allowing a gap to be opened beyond it, while a gap opens after dead zone depletes in the non-zero DZRV model.}
\label{fig:radialpe}
\end{figure*}

% figure 7
\begin{figure*}
\centering
\includegraphics[scale=0.9]{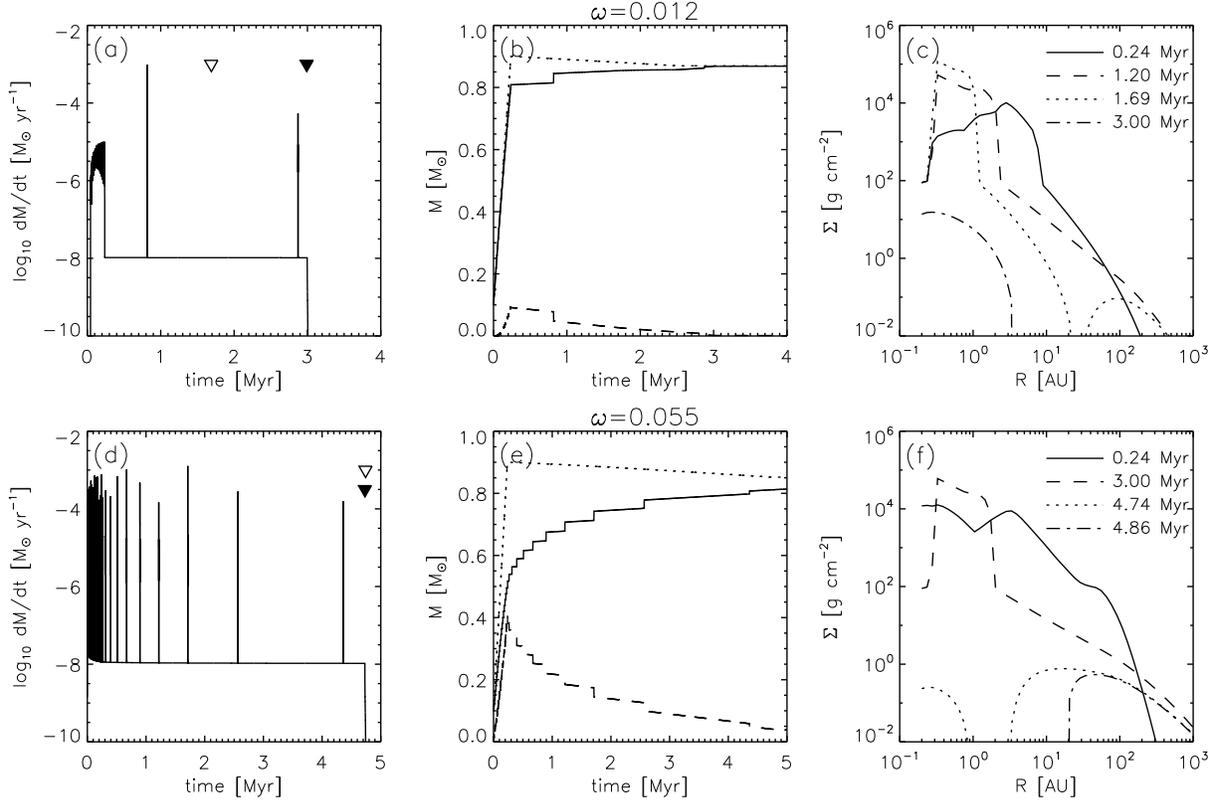}
\caption{Results of photoevaporating models with zero DZRV are presented. (left) Mass accretion rates against time. (middle) Mass of the central star + disk (dotted curve), mass of the central star (solid curves), and mass of the disk (dashed curves). (right) Density distributions at selected times. Upper panels show results of a slowly rotating case ($\omega=0.012$), while lower panel show results of a rapidly rotating case ($\omega=0.055$). In (a) and (d), the open triangles indicate the time when a gap is opened, while the solid triangles present the time when dead zone is depleted.}
\label{fig:evolpew_z}
\end{figure*}

% figure 8
\begin{figure*}
\centering
\includegraphics[scale=0.9]{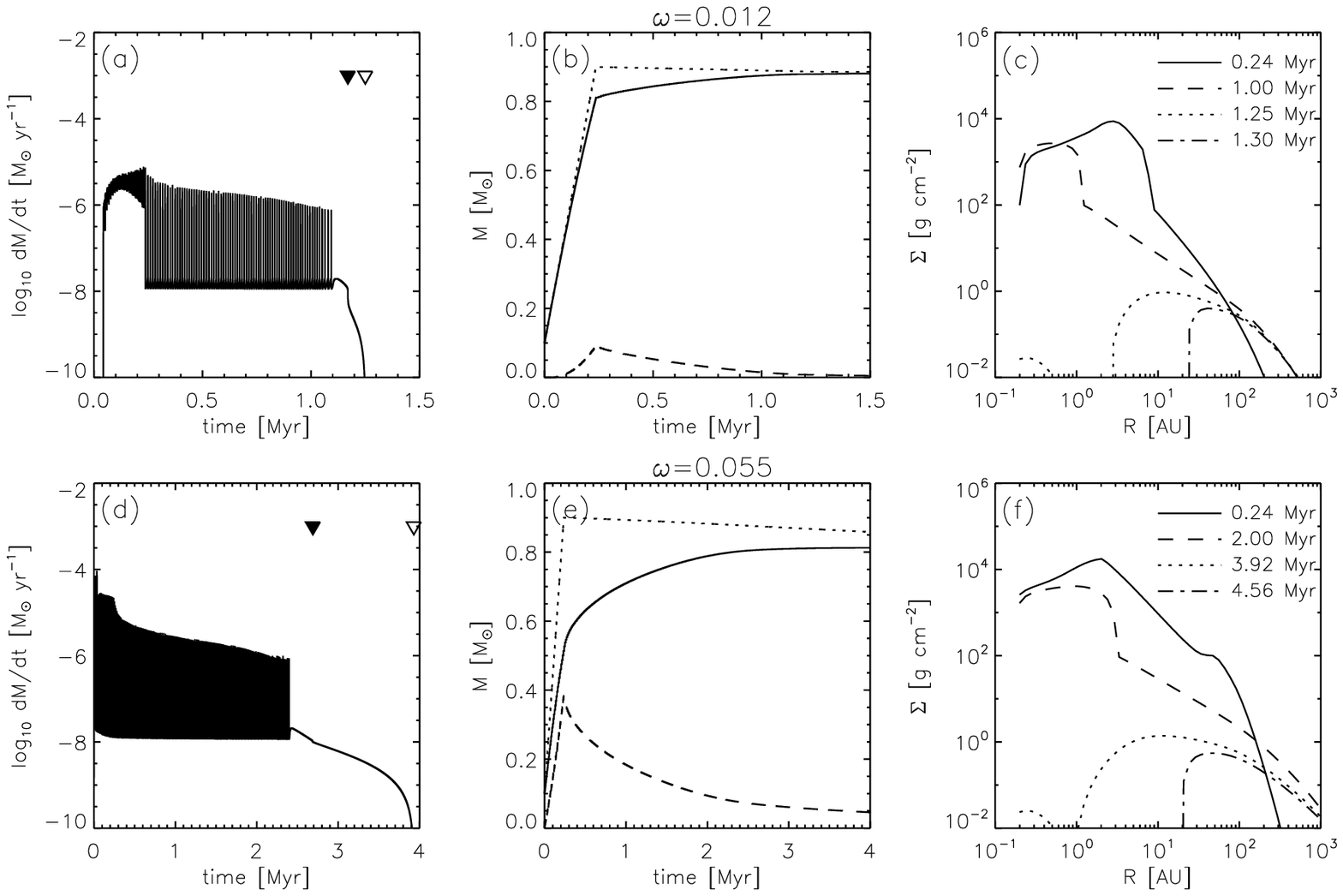}
\caption{Same as Figure \ref{fig:evolpew_z} but results with non-zero DZRV.}
\label{fig:evolpew_nz}
\end{figure*}

% figure 9
\begin{figure*}
\centering
\includegraphics[scale=0.8]{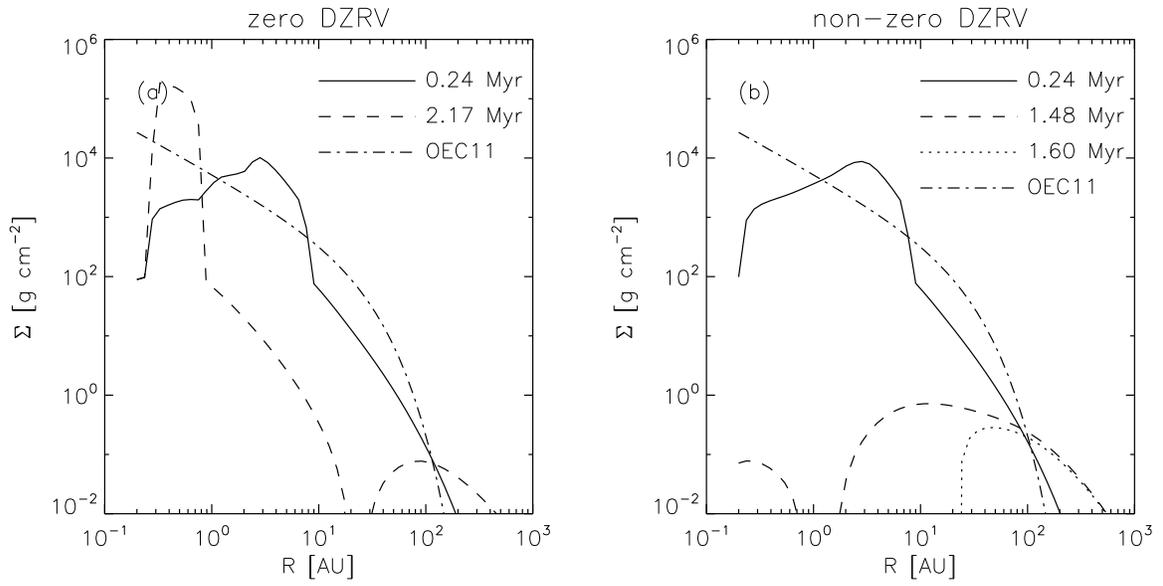}
\caption{Surface density profiles of (a) zero and (b) non-zero DZRV models at the times infall ends (solid curves), a gap opens (dashed curves), and the wall temperature of outer disk drops below 100~K (dotted curve), assuming a comparable photoevaporation rate to OEC11 with $\omega=0.012$. Dashed-dotted curves present the initial surface density distribution of OEC11.}
\label{fig:radialpel}
\end{figure*}

% figure 10
\begin{figure*}
\centering
\includegraphics[scale=1.1]{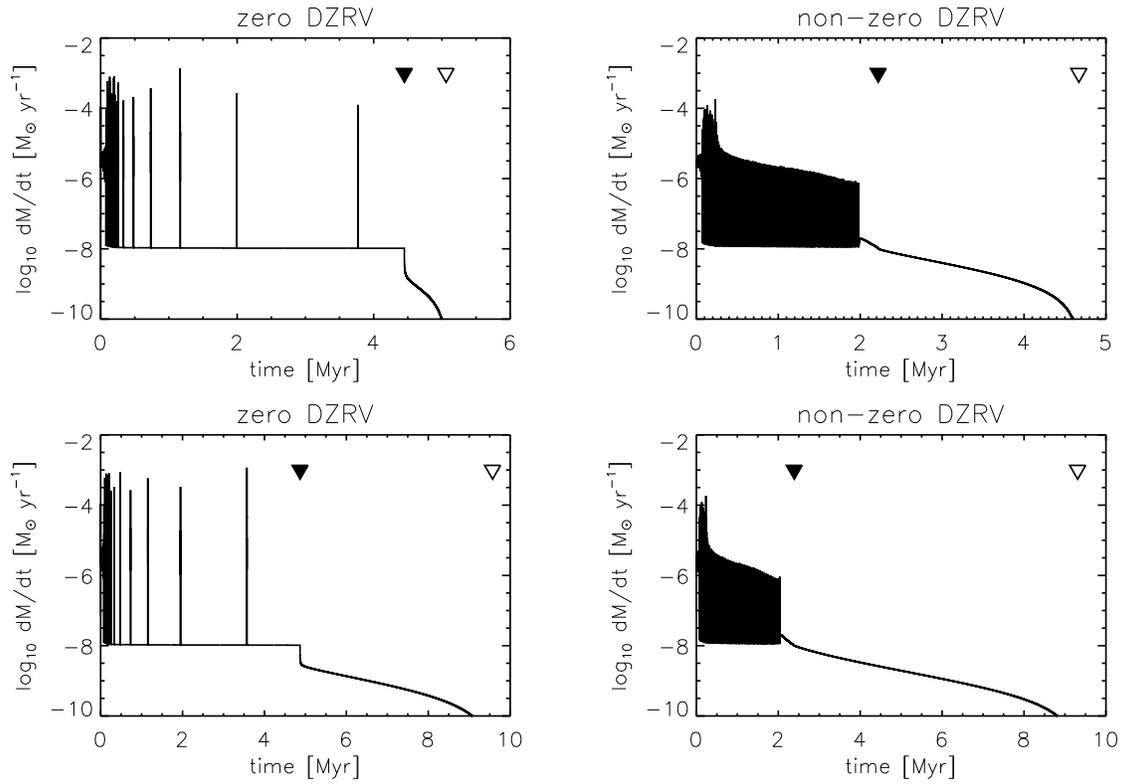}
\caption{Mass accretion rate of photoevaporating models (upper) with $30~\%$ and (lower) with $10~\%$ of the standard photoevaporation rate. The open triangles indicate the time when a gap is opened, while the solid triangles present the time when dead zone is depleted.}
\label{fig:evolpel}
\end{figure*}

% figure 11
\begin{figure*}
\centering
\includegraphics[scale=0.8]{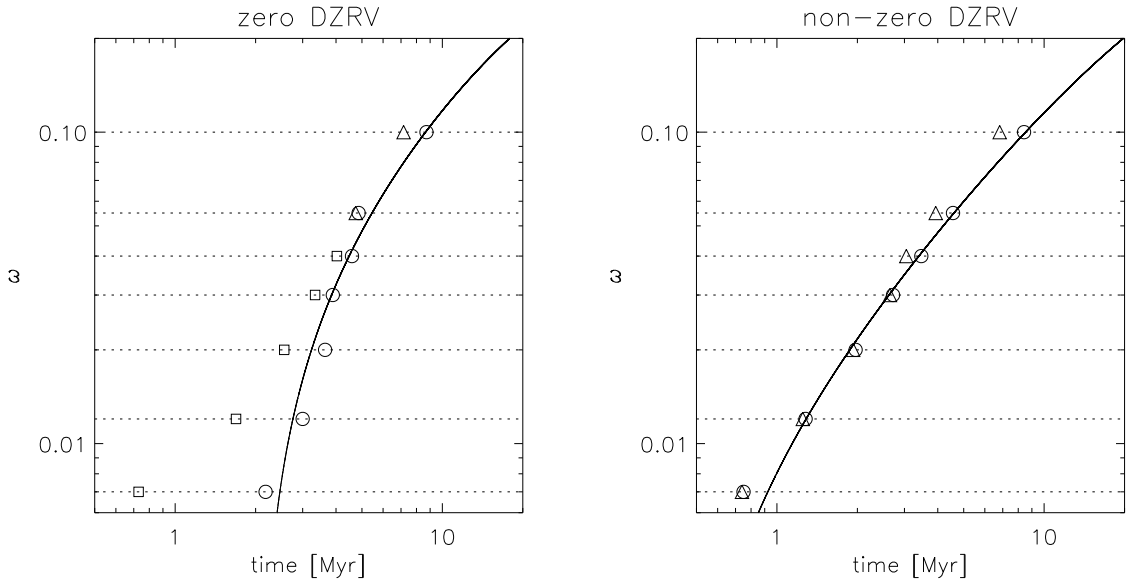}
\caption{A diagram presenting disk phases versus their ages of various disks starting from different initial angular momentum for (a) zero and (b) non-zero DZRV models. Horizontal dotted lines show the path disks evolve on the diagram (from left to right). Circles, triangles, and squares represent evolved disks, accreting transitional disks, and non-accreting transitional disks, respectively. The solid curve is a fit of the evolved disk phase to the initial angular momentum.}
\label{fig:phase}
\end{figure*}

% figure 12
\begin{figure*}
\centering
\includegraphics[scale=0.9]{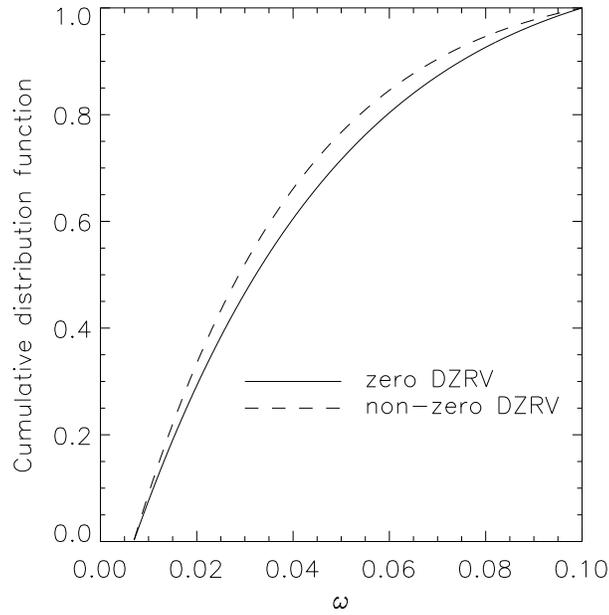}
\caption{Cumulative distribution of initial angular momentum of zero (solid curve) and non-zero (dashed curve) DZRV models.}
\label{fig:am_dist}
\end{figure*}

% figure 13
\begin{figure*}
\centering
\includegraphics[scale=0.9]{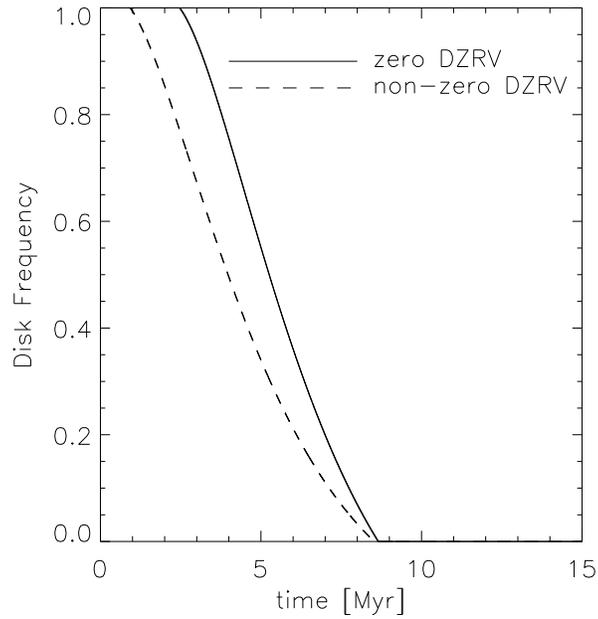}
\caption{Disk frequency of zero (solid curve) and non-zero (dashed curve) DZRV models obtained assuming the initial angular momentum distribution presented in Figure \ref{fig:am_dist}.}
\label{fig:disk_freq}
\end{figure*}

% figure 14
\begin{figure*}
\centering
\includegraphics[scale=0.8]{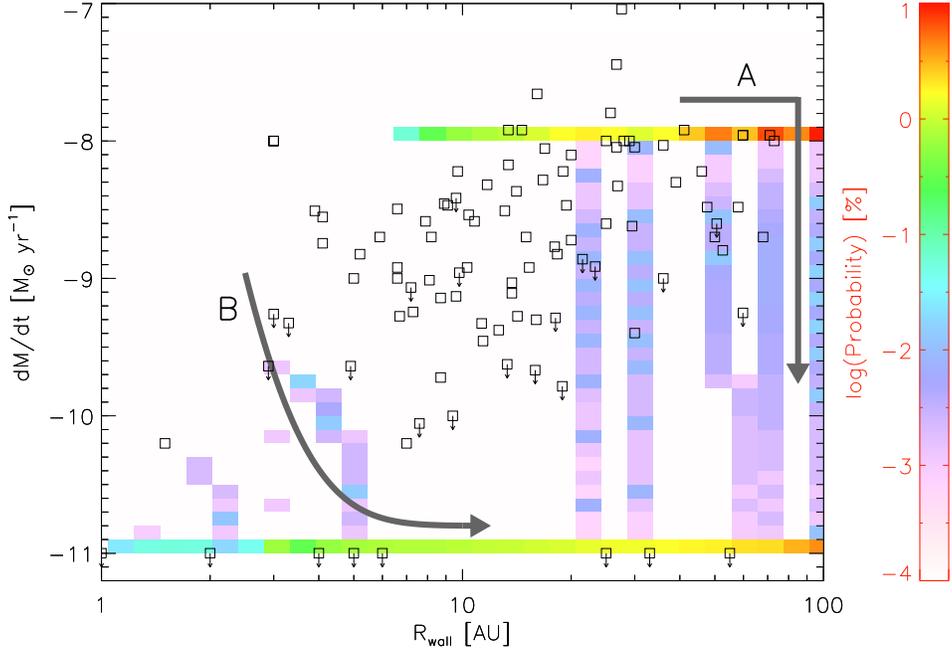}
\caption{The probability distribution of transitional disks on the $\dot{M}-R_{\rm wall}$ plane. The squares present observed transitional disks taken from literatures. The data points with arrows indicate the upper limit of the accretion rate. Objects classified as non-accreting are plotted with an accretion rate of $10^{-11}~\msunyr$. One can see two distinct evolutionary paths; (A) transitional disks with gaps have high accretion rate ($\sim10^{-8}~\msunyr$) initially and move to right before they stop accreting, while (B) transitional disks with inner holes have lower accretion rate ($\lesssim 10^{-9}~\msunyr$) initially, move downward, and then move to right after stop accreting. It should be noted that the transition to the non-accreting phase in both cases is so fast that one can have only a small chance ($<1~\%$) to observe moderately accreting transitional disks.}
\label{fig:mdot_rwall}
\end{figure*}

% figure 15
\begin{figure*}
\centering
\includegraphics[scale=0.8]{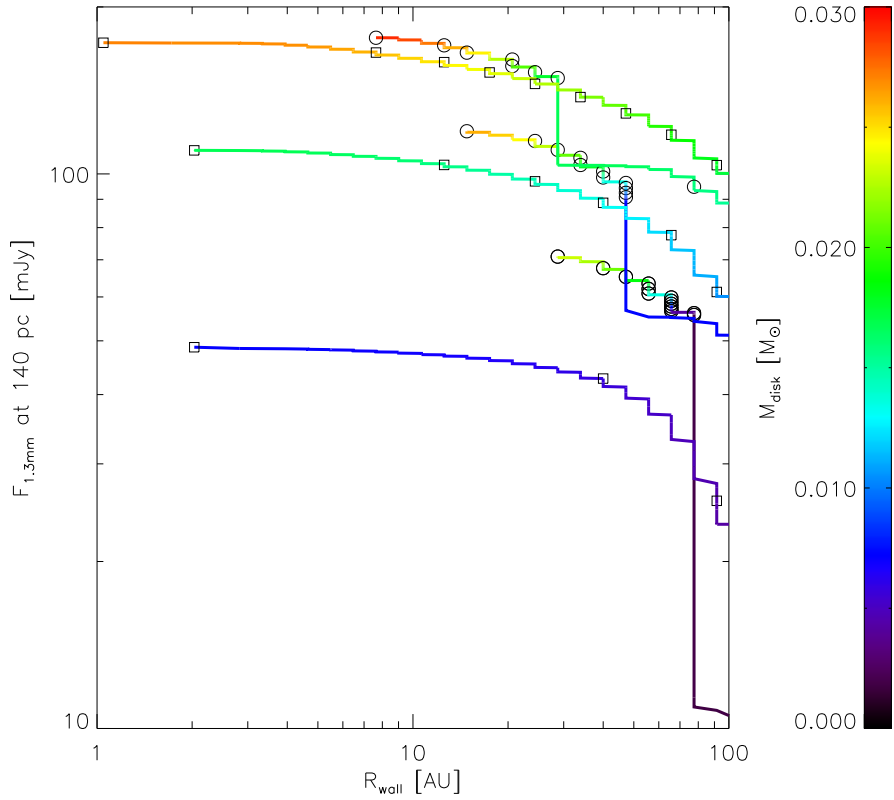}
\caption{Evolution of transitional disks on the ${\rm millimeter~flux} - R_{\rm wall}$ plane with (squares) zero and (circles) non-zero DZRV for three different initial angular momenta; $\omega=0.012$, 0.02, and 0.03 (from bottom to top). Symbols are overplotted every 0.1~Myr and disk masses are color-coded.}
\label{fig:mmflux}
\end{figure*}

% figure 16
\begin{figure*}
\centering
\includegraphics[scale=0.8]{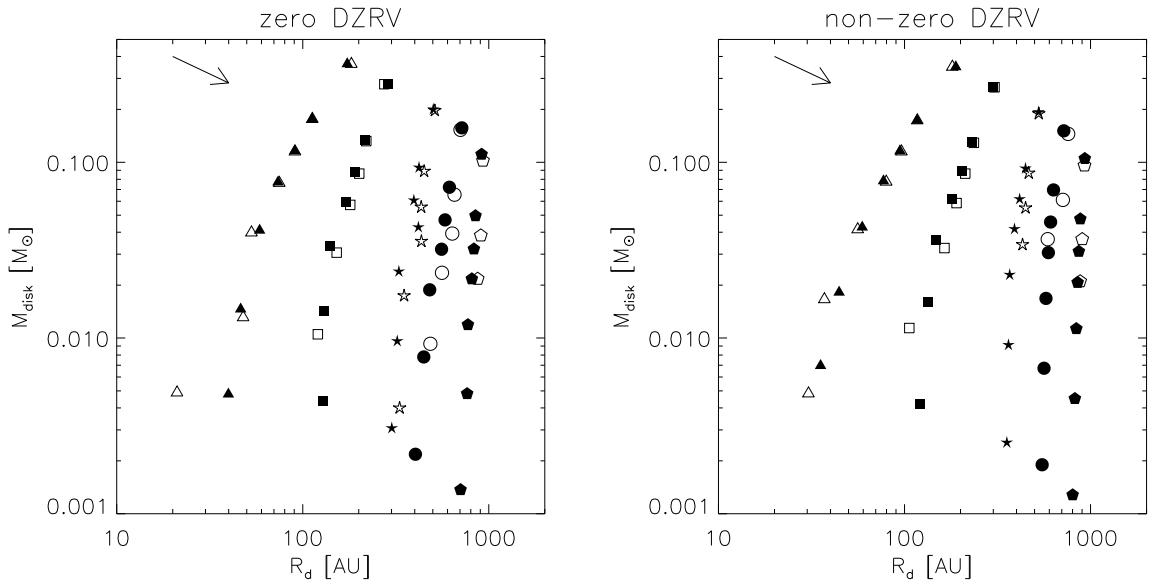}
\caption{The disk masses and characteristic radii of (a) zero and (b) non-zero DZRV models, at 0.5 (triangles), 1 (squares), 2 (stars), 3 (circles), and 5~Myr (pentagons). The filled symbols present non-photoevaporating disks while the opened ones present photoevaporating disks. The arrows indicate the direction of evolution which conserves angular momentum.}
\label{fig:mass_radius}
\end{figure*}

\clearpage

\begin{deluxetable}{ccccccc}
\tablecolumns{7}
\tabletypesize{\small}
\tablecaption{Parameters and results of non-photoevaporating models\label{table:results_wope}}
\tablewidth{0pt}
\tablehead{
\colhead{$\alpha_{\rm rd}$} & 
\colhead{$\omega$} &
\colhead{$R_{\rm c,max}$} &
\colhead{$t_{\rm dead}$} &
\colhead{$M_*$\tablenotemark{a}} &
\colhead{$M_{\rm disk}$\tablenotemark{a}} &
\colhead{$\dot{M}_{\rm acc}(t=10~{\rm Myr})$} \\
\colhead{} & 
\colhead{} &
\colhead{(AU)} &
\colhead{(Myr)} &
\colhead{($M_{\odot}$)} &
\colhead{($M_{\odot}$)} &
\colhead{($M_{\odot}~{\rm yr}^{-1}$)} 
 }
\startdata
zero & 0.007 & 1.0 &2.59 & 0.87/0.90/0.90 & 0.034/0.003/0.001 & $7.39\times 10^{-11}$  \\
zero & 0.012 & 3.1 &3.59 & 0.81/0.89/0.90 & 0.091/0.007/0.003 & $2.61\times 10^{-10}$  \\
zero & 0.02 & 10.0 &4.52 & 0.71/0.89/0.89 & 0.193/0.013/0.006 & $6.49\times 10^{-10}$ \\
zero & 0.03 & 25.2 &5.19 & 0.64/0.88/0.89 & 0.260/0.021/0.012 & $1.19\times 10^{-9}$  \\
zero & 0.04 & 50.4 &5.71 & 0.57/0.87/0.88 & 0.328/0.029/0.017 & $1.78\times 10^{-9}$ \\
zero & 0.055 & 108.2 &6.53  & 0.50/0.86/0.87 & 0.404/0.040/0.027 & $2.77\times 10^{-9}$ \\
zero & 0.1 & 529.5 &9.06  & 0.33/0.83/0.84 & 0.570/0.068/0.061 & $6.31\times 10^{-9}$ \\
non-zero & 0.007 & 1.0 & 0.83  & 0.87/0.89/0.90 & 0.032/0.005/0.001 & $7.05\times 10^{-11}$  \\
non-zero & 0.012 & 3.1 & 1.39 & 0.81/0.89/0.90 & 0.089/0.012/0.002 & $2.49\times 10^{-10}$ \\
non-zero & 0.02 & 9.8 & 1.94 & 0.72/0.88/0.89 & 0.175/0.023/0.006 & $6.23\times 10^{-10}$ \\
non-zero & 0.03 & 25.1 & 2.50 & 0.65/0.87/0.89 & 0.249/0.035/0.011 & $1.15\times 10^{-9}$ \\
non-zero & 0.04 & 48.9 & 3.08 & 0.59/0.86/0.88 & 0.313/0.045/0.017 & $1.74\times 10^{-9}$ \\
non-zero & 0.055 & 105.3 & 3.92 & 0.51/0.84/0.87 & 0.386/0.057/0.026 & $2.67\times 10^{-9}$  \\
non-zero & 0.1 & 504.1 & 6.45 & 0.35/0.81/0.84 & 0.547/0.087/0.059 & $5.99\times 10^{-9}$  \\
\enddata
\tablenotetext{a}{Masses at the end of infall, at the time deadzone depletes, and at 10~Myr.}
\end{deluxetable}

\begin{deluxetable}{cccccccccccc}
\tablecolumns{12}
\tabletypesize{\tiny}
\tablecaption{Parameters and results of photoevaporating models\label{table:results_wpe}}
\tablewidth{0pt}
\tablehead{
\colhead{$\alpha_{\rm rd}$} & 
\colhead{PE rate} &
\colhead{$\omega$} &
\colhead{$t_{\rm dead}$} &
\colhead{$t_{\rm gap}$} &
\colhead{$t_{\rm acc}$} &
\colhead{$t_{\rm evol}$} &
\colhead{$R_{\rm gap}$} &
\colhead{$M_*$\tablenotemark{a}} &
\colhead{$M_{\rm disk}$\tablenotemark{a}} &
\colhead{$\log j$} &
\colhead{Disk phase\tablenotemark{b}} \\
\colhead{} & 
\colhead{} & 
\colhead{} &
\colhead{(Myr)} &
\colhead{(Myr)} &
\colhead{(Myr)} &
\colhead{(Myr)} &
\colhead{(AU)} &
\colhead{($M_{\odot}$)} &
\colhead{($M_{\odot}$)} &
\colhead{$({\rm cm}^2~{\rm s}^{-1})$} &
\colhead{} 
 }
\startdata
zero & 1 &0.007 & 2.18 & 0.73 & 2.19 & 2.19 & 55.7 & 0.87/0.88 & 0.034/0.022 & 17.2 & F$\to$AT$\to$NT  \\
zero & 1 &0.012 & 3.00 & 1.69 & 3.01 & 3.01 & 28.7 & 0.81/0.86 & 0.091/0.027  &  17.8& F$\to$AT$\to$NT  \\
zero & 1 &0.02 & 3.63 & 2.56 & 3.64 & 3.64 & 14.8 & 0.71/0.85 & 0.193/0.026 & 18.0& F$\to$AT$\to$NT   \\
zero & 1 &0.03 & 3.88 & 3.26 & 3.89 & 3.89 & 7.6 & 0.64/0.84 & 0.260/0.029 & 18.6 & F$\to$AT$\to$NT\\
zero & 1 &0.04 & 4.58 & 3.89 & 4.59 & 4.59 & 6.5 & 0.57/0.82 & 0.328/0.038 & 19.1 & F$\to$AT$\to$NT \\
zero & 1 &0.055 & 4.73 & 4.74 & 4.74 & 4.86 & 2.8 & 0.50/0.81 & 0.404/0.041& 19.7 & F$\to$NT \\
zero & 1 &0.1 & 6.35 & 7.15 & 7.15 & 8.71 & 0.6 & 0.33/0.77 & 0.570/0.071& 20.3 & F$\to$NT  \\
non-zero & 1& 0.007 & 0.71 & 0.74 & 0.74 & 0.75 & 3.3 & 0.87/0.89 & 0.032/0.001 & 17.4 & F$\to$NT \\
non-zero & 1& 0.012 & 1.17 & 1.25 & 1.25 & 1.30 & 2.0 & 0.81/0.88 & 0.089/0.007 & 17.9 & F$\to$NT \\
non-zero & 1& 0.02 & 1.60 & 1.83 & 1.83 & 1.97 & 1.5 & 0.72/0.86 & 0.175/0.017 & 18.2 & F$\to$NT  \\
non-zero & 1& 0.03 & 1.94 & 2.44 & 2.44 & 2.72 & 1.0 & 0.65/0.85 & 0.249/0.027& 18.7 & F$\to$NT  \\
non-zero & 1& 0.04 & 2.23 & 3.05 & 3.05 & 3.47 & 1.0 & 0.59/0.83 & 0.313/0.036  & 19.2 & F$\to$NT\\
non-zero & 1& 0.055 & 2.69 & 3.92 & 3.92 & 4.56 & 0.9 & 0.51/0.81 & 0.386/0.047 & 19.7 & F$\to$NT  \\
non-zero & 1 &0.1 & 4.21 & 6.81 & 6.81 & 8.41 & 0.8 & 0.35/0.77 & 0.547/0.071& 20.3  & F$\to$NT \\
zero & 0.6 (OEC11) &0.012 & 2.98 & 2.17 & 3.00 & 3.00 & 24.3 & 0.81/0.86 & 0.091/0.025&  17.8 & F$\to$AT$\to$NT  \\
non-zero & 0.6 (OEC11) & 0.012 & 1.24 & 1.48 & 1.48 & 1.60 & 1.5 & 0.81/0.88 & 0.089/0.007 & 17.9 & F$\to$NT \\
zero & 0.3 & 0.03 & 4.45 & 5.06 & 5.06 & 5.71 & 0.9 & 0.64/0.86 & 0.260/0.017&  18.6 & F$\to$NT  \\
non-zero & 0.3 & 0.03 & 2.22 & 4.67 & 4.67 & 5.41 & 0.9 & 0.81/0.87 & 0.089/0.018 & 18.7 & F$\to$NT \\
zero & 0.1 & 0.03 & 4.87 & 9.57 & 9.57 & 13.72 & 0.8 & 0.64/0.88 & 0.260/0.008 &  18.6 & F$\to$NT  \\
non-zero & 0.1 & 0.03 & 2.39 & 9.31 & 9.31 & 13.49 & 0.8 & 0.81/0.88 & 0.089/0.008 & 18.7 & F$\to$NT \\
\enddata
\tablenotetext{a}{Masses at the end of infall and at the gap opening.}
\tablenotetext{b}{Diks phases indicated as F, AT, and NT denote full disks, accreting transitional disks, and non-accreting transitional disks, respectively.}
\end{deluxetable}

\end{document}